\documentclass[conference]{IEEEtran}
\IEEEoverridecommandlockouts
\usepackage{cite}
\usepackage{amsmath,amssymb,amsfonts}
\usepackage{algorithmic}
\usepackage{graphicx}
\usepackage{textcomp}
\usepackage{xcolor}

\usepackage{color}
\usepackage{amsmath}
\usepackage{tcolorbox}
\usepackage{cite}
\usepackage{multirow}
\usepackage{amsmath,amssymb,amsfonts}
\usepackage{algorithmic}
\usepackage{graphicx}
\usepackage{textcomp}
\usepackage{booktabs}
\usepackage{subfigure}
\usepackage{url}
\usepackage{balance}
\usepackage{todonotes}
\usepackage{array,booktabs}

\usepackage{algorithmic}
\usepackage[ruled, linesnumbered]{algorithm2e}
\usepackage{algorithmic,algorithm2e,float}

\newcommand{\refdiff}{{\sc refdiff}}

\def\BibTeX{{\rm B\kern-.05em{\sc i\kern-.025em b}\kern-.08em
    T\kern-.1667em\lower.7ex\hbox{E}\kern-.125emX}}
\begin{document}

%\author{\IEEEauthorblockN{Aline Brito\IEEEauthorrefmark{1},Andre Hora\IEEEauthorrefmark{1}, Marco Tulio Valente\IEEEauthorrefmark{1}}
%	\IEEEauthorblockA{\IEEEauthorrefmark{1}
%	ASERG Group, Department of Computer Science (DCC)\\  
%	Federal University of Minas Gerais (UFMG), Brazil\\
%	\{alinebrito, andrehora, mtov\}@dcc.ufmg.br
%	}
%}

\title{Refactoring Graphs:\\ Assessing Refactoring over Time\\
% \title{Assessing Refactoring Over Time\\
%{
%	\footnotesize \textsuperscript{*}Note: Sub-titles are not captured in Xplore and
}

 \author{\IEEEauthorblockN{
 Aline Brito, Andre Hora, Marco Tulio Valente}
 	\IEEEauthorblockA{ASERG Group, Department of Computer Science (DCC), Federal University of Minas Gerais, Brazil\\
 	\{alinebrito, andrehora, mtov\}@dcc.ufmg.br
 	}
}

%\author{\IEEEauthorblockN{
%[Omitted due to DBR]}
%	\IEEEauthorblockA{\hspace{1cm}\\
%	\hspace{1cm}
%	}
%}

\maketitle

\begin{abstract}
Refactoring is an essential activity during software evolution. 
Frequently, practitioners rely on such transformations to improve source code maintainability and quality.
As a consequence, this process may produce new source code entities or change the structure of existing ones.
Sometimes, the transformations are atomic, i.e., performed in a single commit. In other cases, they generate sequences of modifications performed over time. To study and reason about refactorings over time, in this paper, we propose a novel concept called refactoring graphs and provide an algorithm to build such graphs.
Then, we investigate the history of 10 popular open-source Java-based projects.
After eliminating trivial graphs, we characterize a large sample of 1,150 refactoring graphs, providing quantitative data on their size,
commits, age, refactoring composition, and developers.
We conclude by discussing applications and implications of refactoring graphs, for example, to improve code comprehension, detect refactoring patterns, and support software evolution studies.

\end{abstract}


\section{Introduction}
\label{section:introduction}

Refactoring is a key activity to preserve and evolve the
internal design of software systems. Due to the
importance of the practice in modern software 
development, there is a large body of papers 
and studies about refactoring, 
shedding light on aspects such as usage of 
refactoring engines~\cite{Murphy-Hill:ICSE:2009,Negara:2013:ECOOP}, documentation 
of refactorings using commit messages~\cite{Murphy-Hill:ICSE:2009},
motivations for performing
refactorings~\cite{danilo:fse2016:WhyWeRefactor,Mazinanian:2017:OOPSLA,Tsantalis:2013:CASCON}, benefits
and challenges of refactoring~\cite{Kim:2012:FSE, Kim:2014:TSE},
among many others.

However, \textit{time} seems to be an underinvestigated dimension
in refactoring studies. The notable exception are
studies on refactoring tactics, particularly
on repeated refactoring operations, often
called {\em batch} refactorings. For example, Murphy-Hill \textit{et al.}~\cite{Murphy-Hill:ICSE:2009} define batch refactorings as
operations that execute within 60 seconds 
of each another. They report that 40\% of refactorings performed 
using a refactoring tool occur in
batches, i.e., programmers repeat refactorings. But the authors also mention that
``\emph{the main limitation of [our] analysis is that, while we
wished to measure how often several 
related refactorings are performed in sequence, 
we instead used a 60-second heuristic}''.
Bibiano~\textit{et al.}~\cite{Bibiano:esem2019:BatchRefactoring} investigate the characteristics and impact of batch refactorings on code elements affected by smells. 
The authors rely on a heuristic to retrieve batches~\cite{cedrimRego:2018}, which groups refactorings performed by the same author in a single code element. 
Thus, their heuristic focus on single methods or classes, most of the cases resulting in batches with a single commit (93\%).

Interestingly, in his seminal book on
refactoring~\cite{Fowler:1999}, Fowler dedicates a chapter---co-authored with Kent Beck---to {\em big refactorings}. They claim
that when studied individually refactorings do
not provide a whole picture of the ``game'' played by developers when improving software design, i.e., ``\emph{refactorings take
time [to be concluded]}''. However, to our knowledge, refactorings performed over
long time windows are not deeply studied by the literature.

Therefore, we propose and evaluate a novel concept, called
{\bf refactoring graphs}, to study and reason about refactoring activities
over time. In such graphs, the nodes are methods
and the edges represent refactoring operations. For example,
suppose that a method $ f\!oo()$ is renamed to $bar()$. This
operation is represented by two nodes, $f\!oo()$ and
$bar()$, and one edge connecting them. After this
first refactoring,
suppose that a method $qux()$ is extracted
from $bar()$. As a result, an edge connecting $bar()$ to
a new node, representing $qux()$, is also added to
the graph. Furthermore, 
refactoring graphs do not impose
time constraints between
the represented refactoring operations.
In our example, the extract operation,
for instance, can be performed months
after the rename. Finally, refactoring
graphs may also express refactorings performed by
different developers. In our example, the rename can be
performed by $d_1$ and the extract by
another developer $d_2$.

We formalize an algorithm to build
refactoring graphs and use it to to
extract graphs for 10 well-known
and popular open-source Java-based
projects. 
Our goal is to characterize refactoring subgraphs to better understand this scenario. Thus,
after removing refactoring
graphs coming from a single
commit (since our goal is to investigate
refactorings over time), we answer
five research questions about the
following
properties:
%of refactoring graphs:

\begin{itemize}

\item Size (RQ1): most refactoring graphs have at most four nodes (85\%) and three edges (83\%). However, we also found graphs with 57 nodes and 61 edges. 

\item Commits (RQ2): Most refactoring subgraphs are generated from two or three commits (95\%).

\item Age (RQ3): The age of the refactoring subgraphs ranges from a few days to weeks or even months. For instance, 67\% of the subgraphs have more than one month.

\item Refactoring composition (RQ4): Most refactoring subgraphs include more than one refactoring type (72\%).

\item Developers (RQ5): Most refactoring subgraphs are created by a single developer (60\%). However, a relevant amount (40\%) is created by multiple developers.

\end{itemize}

Our main contributions are threefold. \emph{First}, we propose and formalize the notion of refactoring graphs, which can be used to study and reason about refactorings performed over any time window by multiple developers. \emph{Second}, we reveal several properties of a large sample of 1,150 refactoring graphs extracted for 10
real software projects. \emph{Third}, we discuss several applications and implications of refactoring graphs to expand current refactoring tools, improve code comprehension, detect refactoring patterns, and support software evolution studies.\\[-0.2cm]

\noindent{\em Structure:} Section \ref{section:refactoring-graph}  defines our concept of refactoring graphs.  Section \ref{section:studyDesing} describes the design of our study, while Section \ref{section:results} shows the results. 
Section \ref{sec:example} shows an example of a large refactoring subgraph.
We discuss the key applications and implications in Section \ref{section:discussion}.  Section \ref{section:threatsValidity} states threats to validity and Section \ref{section:relatedWork} presents related work. Finally, we conclude the paper in Section \ref{section:conclusion}.

\section{Refactoring Graphs}
\label{section:refactoring-graph}

% https://www.sciencedirect.com/topics/computer-science/connected-subgraph
%https://en.wikipedia.org/wiki/Component_(graph_theory)
%https://xlinux.nist.gov/dads/HTML/subgraph.html

\begin{figure}[!t]
	\centering
	%\vspace{0.3cm}
    \includegraphics[width=0.49\textwidth]{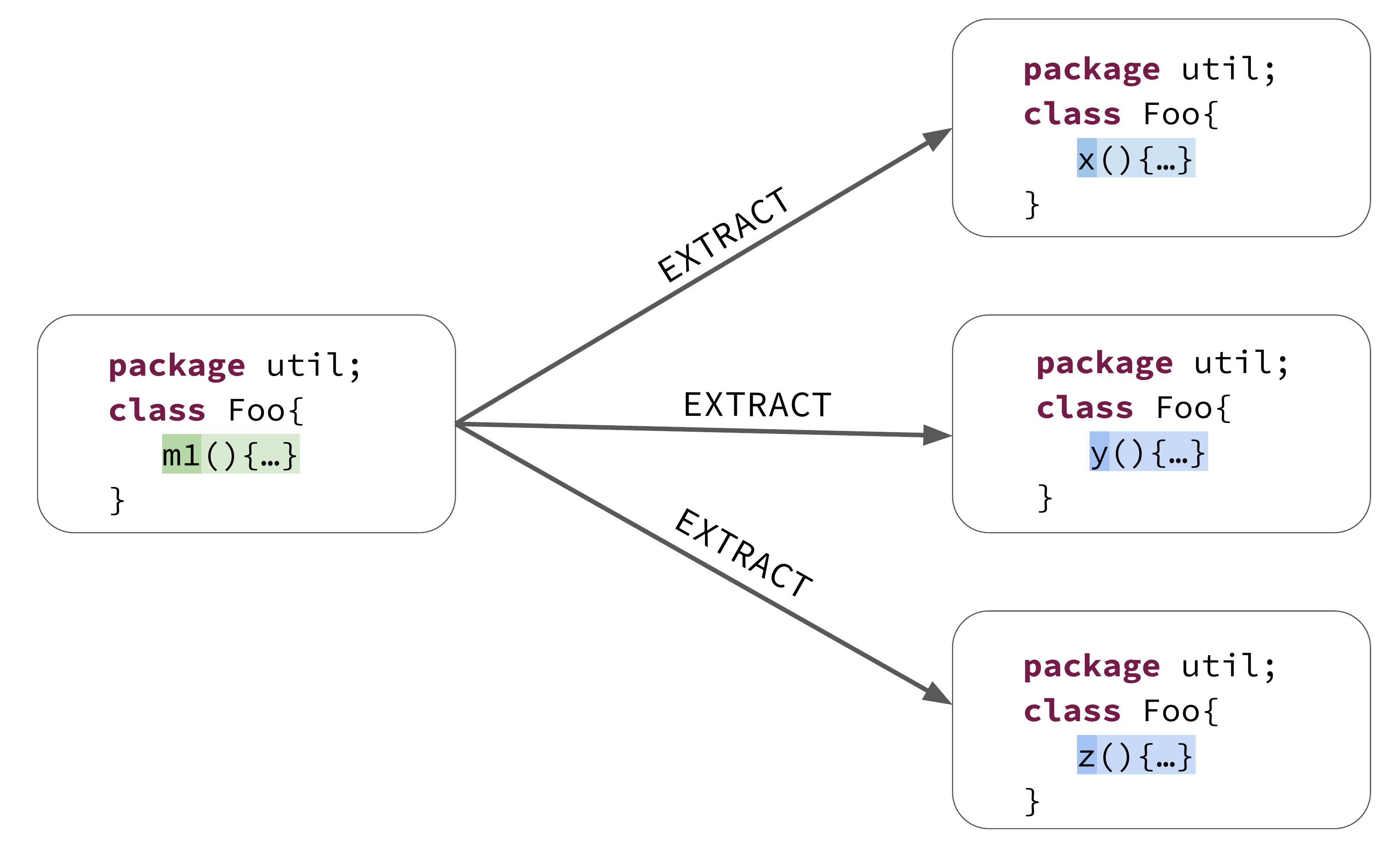}
	\caption{Refactoring subgraph produced by only one developer}
	\label{fig:example_refactoring_graph_1_trivial}
\end{figure}

A \textit{refactoring graph} $G$ is a set of disconnected subgraphs $G' = (V', E')$.
Each $G'$ is called a \textit{refactoring subgraph}, with a set of vertices $V'$ and a set of directed edges $E'$. 
In this way, the history of a  software system includes a set of refactoring subgraphs.
In refactoring (sub)-graphs, the vertices are the full signature of  methods.
For instance, we labeled a method $m()$  in class $F\!oo$ and package $util$ as $util.F\!oo\#m()$.
Finally, the edge indicates the refactoring type (e.g., \textit{move method}) and it also includes meta-data about the operation (e.g., author name and date).

\begin{figure}[!t]
	\centering
    \includegraphics[width=0.49\textwidth]{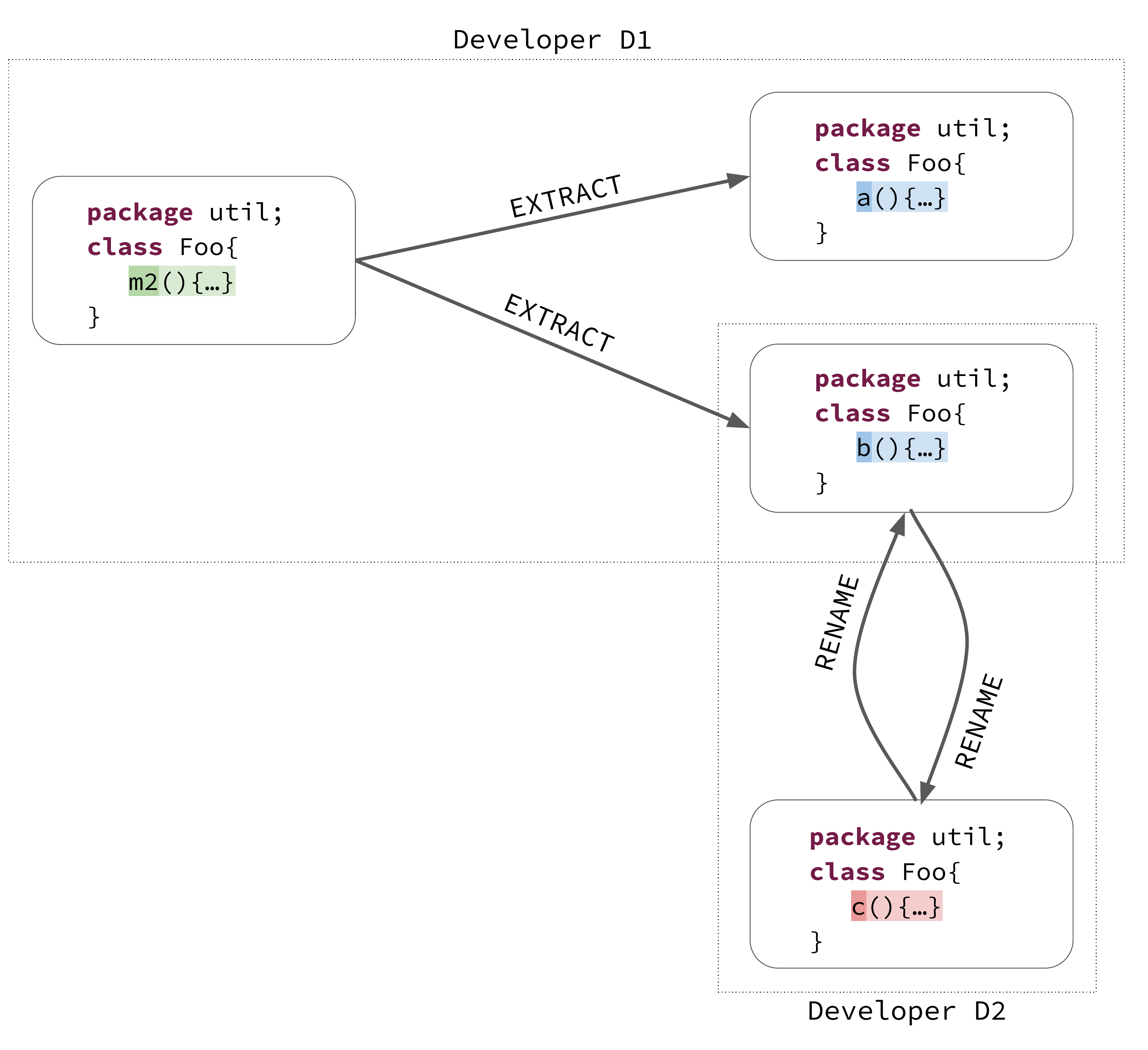}
	\caption{Refactoring subgraph over time}
	\label{fig:example_refactoring_graph_3_complexo}
\end{figure}

Figure \ref{fig:example_refactoring_graph_1_trivial} shows an example of a \textit{refactoring graph}.
A developer extracted three methods from $m1()$, which are named $x()$,  $y()$, and $z()$. 
The edges refer to the refactoring operation.
It is worth noting that a refactoring graph can include refactorings performed by multiple developers. 
For instance, Figure~\ref{fig:example_refactoring_graph_3_complexo} illustrates a second example, where a developer $D1$ extracted two methods from $m2()$, which are named $a()$ and $b()$. 
Then, a second developer $D2$ renamed $b()$ to $c()$.
After that, a code reviewer might have suggested to keep the original name. Thus,  the developer undoes the latest refactoring,  renaming $c()$ to $b()$  again. 
In this case, the graph contains refactorings performed by two authors. 
Besides,  there is a cycle when the developer reverts the method to the original name. 

%top-10 projetos GitHub
%branch default
%commits recuperados via firt parent
%refdiff v2.0 (suporte para várias linguagens)
%apenas refatorações em nível de metódos

\begin{table*}[!ht]
%\vspace{-0.25cm}
\centering
\caption{selected java projects}
\label{table:selected-projects}
\begin{tabular}{l r r r r r r l}
\toprule
{\bf Project} & {\bf Stars} & {\bf Forks} & {\bf Commits} & {\bf Contributors} & {\bf Java Files} & {\bf Latest Version} & {\bf Description}
\\ \midrule

Elasticsearch & 44,489 & 14,930 & 48,313 & 1,273 & 11,770 & 7.3.2 & Search engine for cloud systems \\ 

RxJava & 40,622 & 6,825  & 5,581 & 237 & 1,666 & 3.0.0-RC3 & Library for asynchronous communications \\

Square Okhttp & 34,484 & 7,521 & 4,273  & 189 & 167 & 4.2.0 & HTTP client\\

%Google Guava &  33,973 & 7,573 & 5,067 & 202 & 3,151 & Google utility libraries \\  

Square Retrofit & 33,801& 6,254  & 1,756 & 129 & 241 & 2.6.2 & HTTP client  \\

Spring Framework & 32,582 & 21,226 & 19,752 & 396 & 7,203 & 5.2.0 & Framework for web aplications \\

Apache Dubbo & 29,353 & 19,256  & 3,639 & 249 & 1,743 & 2.7.3 & RPC framework  \\

MPAndroidChart & 28,647 & 7,424 & 2,018 & 66 & 220 & 3.1.0 & Library to create charts \\

Glide & 27,289 & 5,025  & 2,416 & 102 & 647 & 4.10.0 & Library to load imagens \\

Lottie Android & 26,952 & 4,278 & 1,139 & 76 & 198 & 3.0.7 & Library to parser animations \\

Facebook Fresco & 15,870 & 3,595 & 2,158 & 170 & 985 & 2.0.0 & Library to display images \\

\bottomrule
\end{tabular}        
\end{table*}

\begin{figure}[!b]
	\centering
	\includegraphics[width=3in]{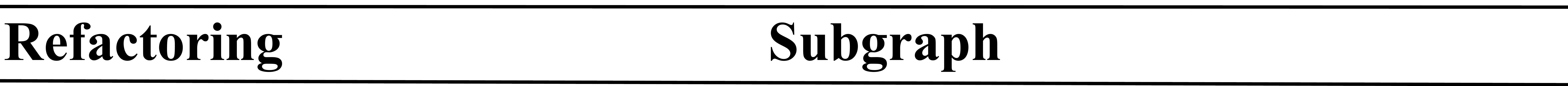}\vspace{0.2cm}
    \includegraphics[width=3in]{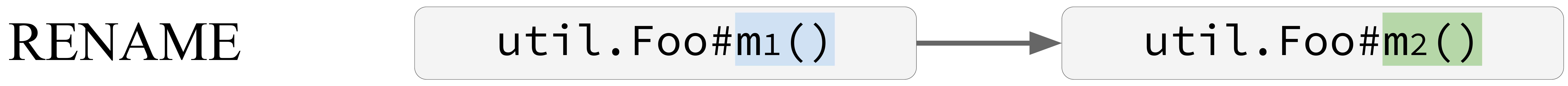}\vspace{0.7cm}
    \includegraphics[width=3in]{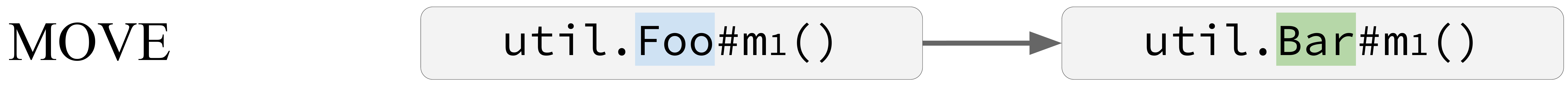}\vspace{0.7cm}
    \includegraphics[width=3in]{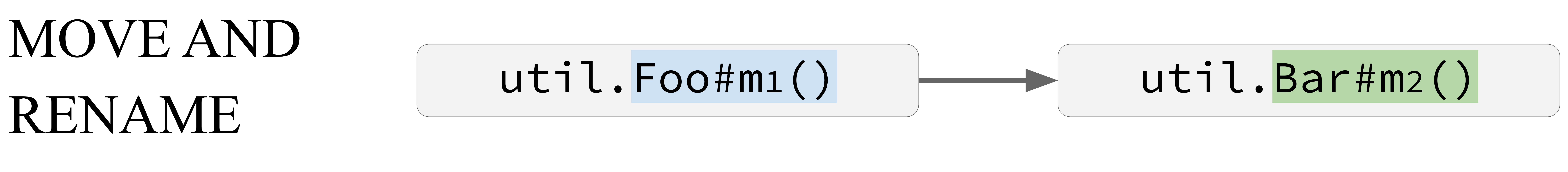}\vspace{0.7cm}
    \includegraphics[width=3in]{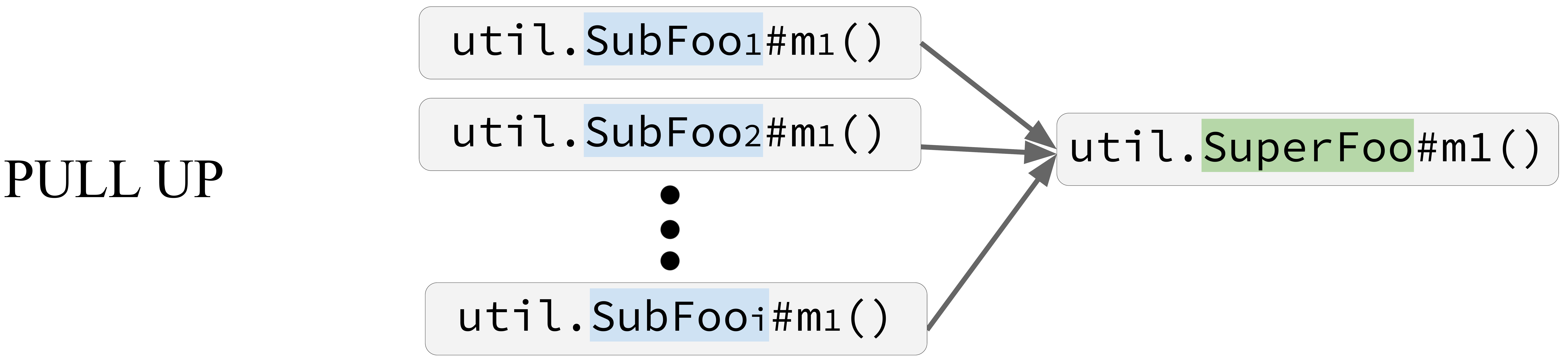}\vspace{0.7cm}
    \includegraphics[width=3in]{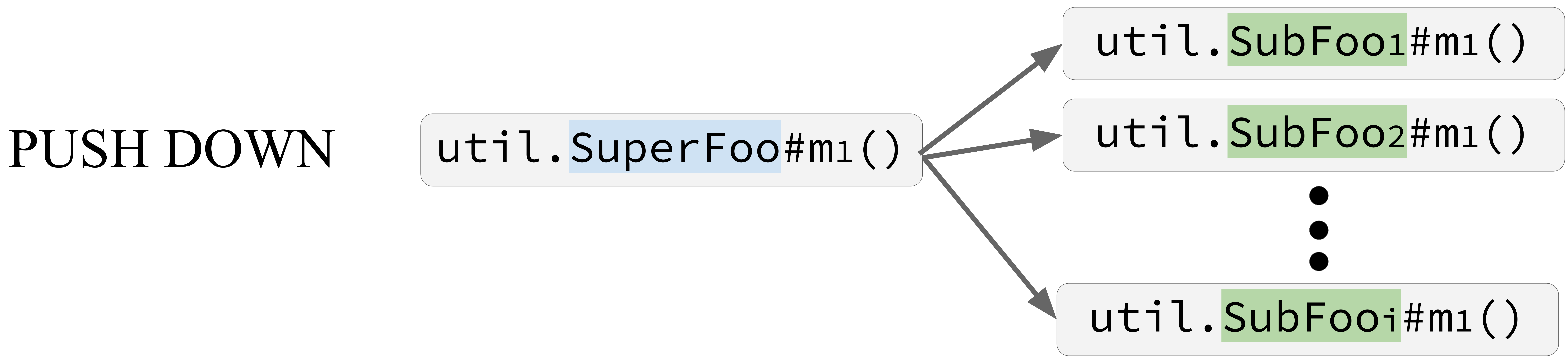}\vspace{0.7cm}
    \includegraphics[width=3in]{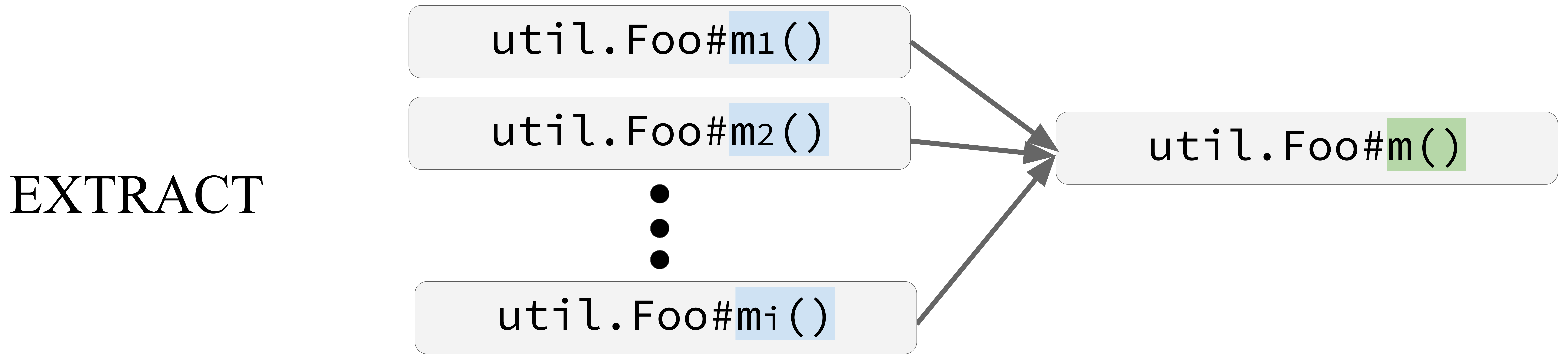}\vspace{0.7cm}
    \includegraphics[width=3in]{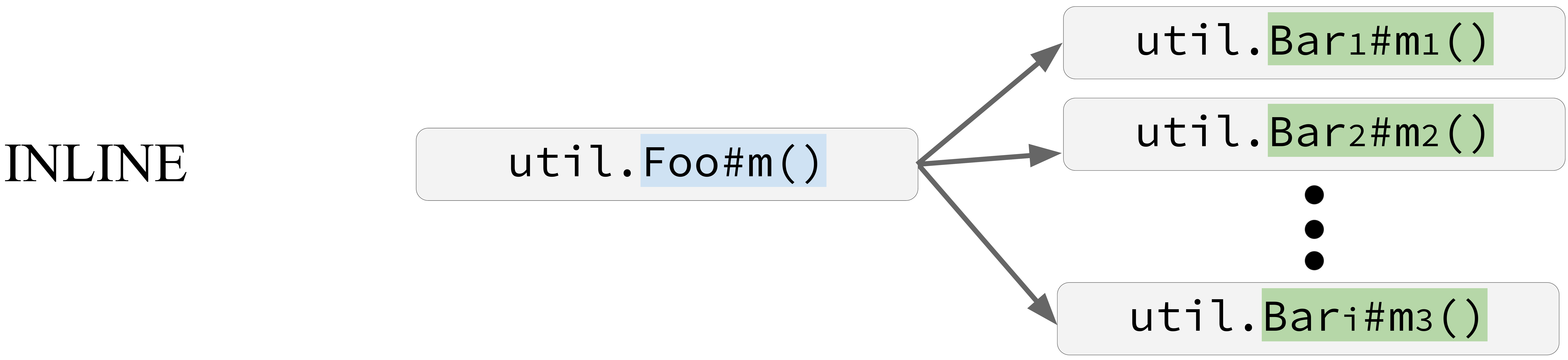}\vspace{0.7cm}
    \includegraphics[width=3in]{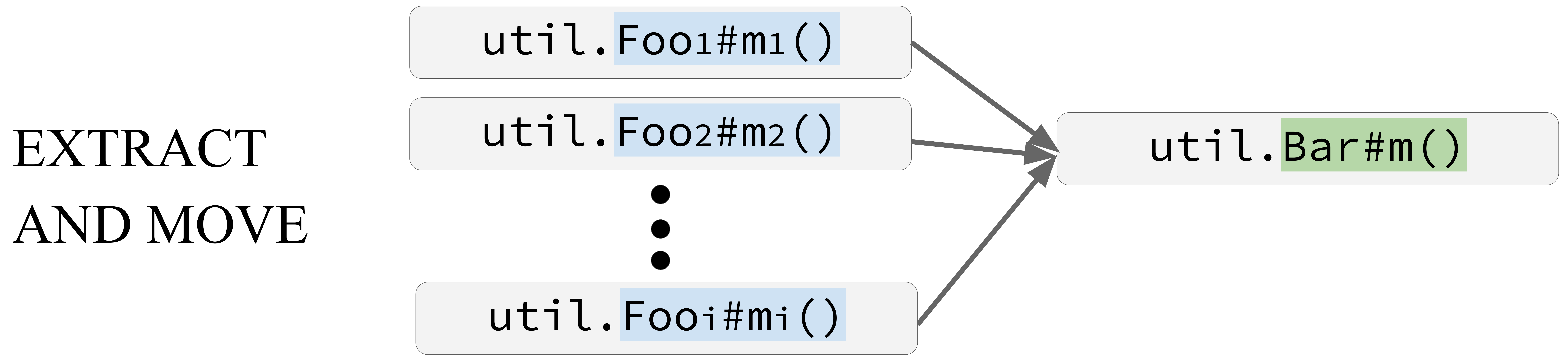} 
    \vspace{0.2cm}
	\caption{Example of refactoring subgraphs}
	\label{fig:graph_refactorings}
\end{figure}

As presented in Figure~\ref{fig:graph_refactorings}, we center our study on eight refactorings at the method level.
\textit{Rename} and \textit{move} are the most trivial operations since they involve just changing the method's signature.
Inheritance-based refactorings comprise the movement of one or more methods to supertypes or subtypes (i.e., \textit{pull up} and \textit{push down}). 
For example, a \textit{pull up} moves methods from subclasses to a superclass.
Extract operations generate new methods in the same class (i.e., they create a new node in our subgraphs). 
It is possible to extract a method $m()$ or multiple methods $m_i$ from a single method $m1()$. 
However, as also illustrated in Figure~\ref{fig:graph_refactorings}, it is possible to extract $m()$ from multiple methods $m_i$. In this case, the extracted code is duplicated in each method $m_i$.
\textit{Inline method} is a dual operation, involving the removal of trivial elements and replacement of the respective calls by their content.
As in the case of \textit{extract}, we can inline a method  $m()$ in multiple methods $m_i$.
Finally, we consider a refactoring called \textit{extract and move} that extracts a method to a distinct class.

\section{Study Design}
\label{section:studyDesing}

\subsection{Selecting Java Projects}

We analyze the characteristics and frequency of refactoring subgraphs in popular software systems. 
We select 10 popular Java projects in terms of stars on GitHub, since stars is a key metric to reveal the popularity of repositories~\cite{hudson:icsme2016:Popularity,jss-2018-github-stars}. 
We also confine our analysis to projects with more than 1K commits and more than 100 Java files to avoid young and small systems.
Table \ref{table:selected-projects} describes the selected projects, including basic information, such as number of stars, commits, files, contributors, latest version, and description.
These projects cover distinct domains, including web development systems and media processing libraries, for example.
The most popular project is {\sffamily \small Elasticsearch} (44,489 stars). 
The number of forks ranges from  3,595 ({\sffamily \small Facebook Fresco}) to  21,226 ({\sffamily \small Spring Framework}).
The number of commits ranges from 1,139 ({\sffamily \small Lottie Android}) to 48,313 ({\sffamily \small Elasticsearch}), while the number of contributors varies from 66 ({\sffamily \small MPAndroidChart}) to 1,273 ({\sffamily \small Elasticsearch}). 
{\sffamily \small Square Okhttp} is the smallest system (167 files); and {\sffamily \small Elasticsearch} is the largest one (11,770 files).  \\[-0.2cm]

\subsection{Detecting Refactoring Operations}

We use \refdiff~\cite{danilo:msr2017:RefDiff}~to detect the refactoring operations 
needed to build refactoring graphs. 
\refdiff~identifies refactorings between two versions of a git-based project. In our study, we focus on well-known refactoring operations detected by \refdiff~at the method level (i.e., rename, move, extract, inline, pull up, and push down, as presented in Figure~\ref{fig:graph_refactorings}).

\refdiff~works by comparing each commit with its previous version in history.
To avoid analyzing commits from temporary branches, we focus on the main branch evolution.  
Particularly, we use the command {\small \sffamily git log --first-parent} to get the list of commits of each project.\footnote{\url{https://git-scm.com/docs/git-log#Documentation/git-log.txt---first-parent}}
Additionally, we remove refactorings in packages with the keywords \textit{test(s)}, \textit{example(s)}, and \textit{sample(s)},  since they are not part of the core system.

\subsection{Building Refactoring Graphs}

As mentioned earlier, we identify refactoring subgraphs over time in 10 systems. 
Algorithm \ref{alg:detectec_refactoring_subgraphs} presents the steps to build refactoring graphs.
The input comprises a list of refactorings, e.g., $util.Foo\#m()$ moved to $util.Bar\#m()$.
First, the algorithm identifies each refactoring $t$ and the two methods involved, $m1$ and $m2$ (line 3).
Then, it creates a directed edge representing this refactoring (line 5). 
%Methods $m1$ and $m2$ are inserted in the graph $DG$ only if there are no vertices to represent them (Line 5). 
Since $V$ and $E$ are sets, each element is represented only one time.
The edges are labeled with refactoring's name $t$. %Besides that, it carries the meta-data about the operation as well as the origin and destination vertices.
The output includes sets of refactoring subgraphs in text format.

%https://oeis.org/wiki/List_of_LaTeX_mathematical_symbols
\begin{algorithm}
    \small
	\SetAlgoLined
	\KwIn{R (list of refactorings from a system S)}
	\KwOut{DG (refactoring graph)}
	\Begin{	
			DG $\gets$ $\emptyset$, V $\gets$ $\emptyset$, E $\gets$ $\emptyset$\

			\For{(m1, m2, t) $\in$ R}{
			
				V $\gets$ V $\cup$ \{$m1, m2$\}\
	
				E $\gets$ E $\cup$  $(m1, m2, t)$\
				
			}
			\textbf{return} (V, E)
			
	}
	\caption{Building refactoring graphs}
	\label{alg:detectec_refactoring_subgraphs}
 \end{algorithm}

 Table~\ref{table:ref-graphs} presents the frequency of refactoring subgraphs in the analyzed systems. Considering all the projects, we detect a total of 8,926 refactoring subgraphs. 
{\sffamily \small Spring Framework} has the highest number of subgraphs (3,104), while {\sffamily \small Square Retrofit} has the lowest amount (169).
Overall, 87.1\% of the refactoring subgraphs comprise a set of operations performed in a single commit.
This ratio varies from 69.2\% ({\sffamily \small Glide}) to 93.8\% ({\sffamily \small Apache Dubbo}).
In contrast, 12.9\% capture refactorings performed in two or more commits.
%In this case, the ratio varies from 6.9\% (XXX) to XXX\% ({\sffamily \small Glide}).
\textbf{In this paper, we assess the 1,150 refactoring subgraphs with number of commits $\ge$ 2, because they are the ones that represent refactoring over time.}

%len(1) são subgrafos que envolvem apenas um commit.
%len >= 2, projetos com pelos menos dois commits distintos
 
 \begin{table}[!ht]
%\vspace{-0.25cm}
\centering
\caption{Frequency of Refactoring subgraphs}
\label{table:ref-graphs}
\begin{tabular}{l r r r r r}
\toprule
\multirow{2}{*}{{\bf Project}} & \multicolumn{5}{c}{{\bf Refactoring Subgraphs}} \\ \cline{2-6}
& All & $len=1$ & \% & $len\ge2$  & \% \\

\midrule

Elasticsearch	&	2,073	&	1,934	&	93.3	&	139	&	6.7	\\
RxJava	&	1,073	&	975	&	90.9    &	98	&	9.1	\\
Square Okhttp	&	635	&	548	&	86.3	&	87	&	13.7 \\
Square Retrofit	&	169	&	135	&	79.9	&	34	&	20.1\\
Spring Framework	&	3,104	&	2,604	&	83.9 &	500	&	16.1	\\
Apache Dubbo	&	483	&	453	&	93.8	&	30	&	6.2	\\
MPAndroidChart	&	454	&	381	&	83.9	&	73	&	16.1	\\
Glide	&	425	&	294	&	69.2	&	131	&	30.8	\\
Lottie Android	&	196	&	173	&	88.3    &	23	&	11.7	\\
Facebook Fresco	&	314	&	279	&	88.9	&	35	&	11.1	\\

\midrule

Total	&	8,926	&	7,776	&	87.1	&	1,150	&	12.9	\\

\bottomrule
\end{tabular}      

\end{table}

\section{Results}
\label{section:results}

%-----------------------------------------
\subsection{(RQ1) What Is the Size of Refactoring Subgraphs?}

\begin{figure}[!ht]
	\centering
    \subfigure{\includegraphics[width=0.45\textwidth]{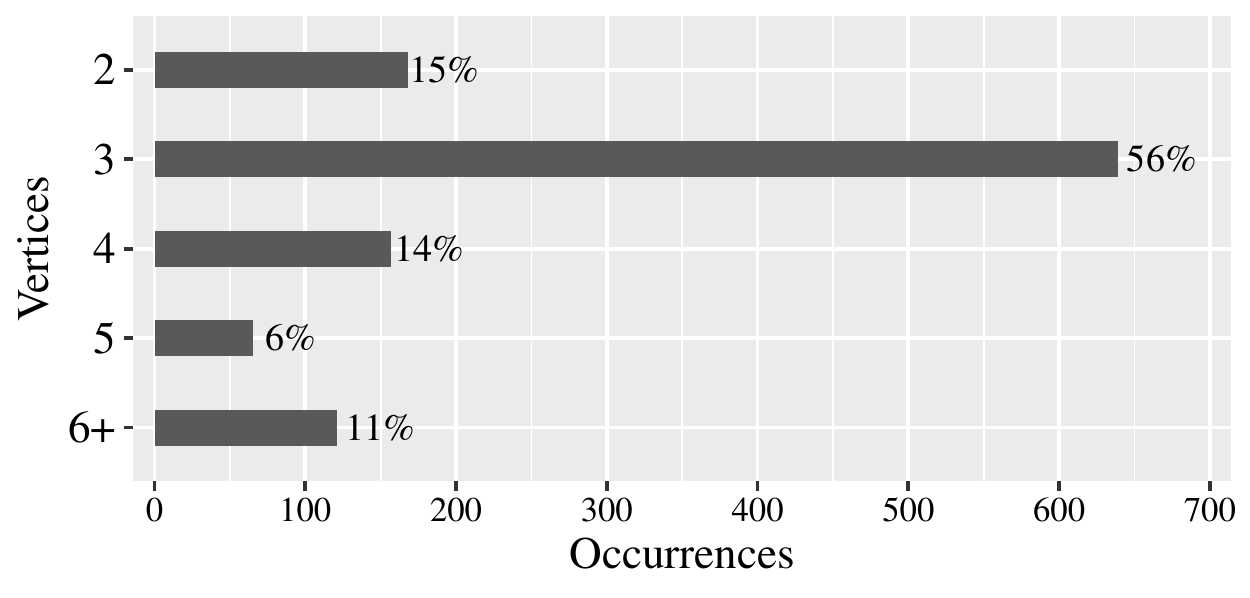}}
	\caption{Number of vertices by refactoring subgraph}
	\label{fig:rq2-size-vertices}
\end{figure}

\begin{figure}[!ht]
	\centering
	%\vspace{0.5cm}
    \subfigure{\includegraphics[width=0.45\textwidth]{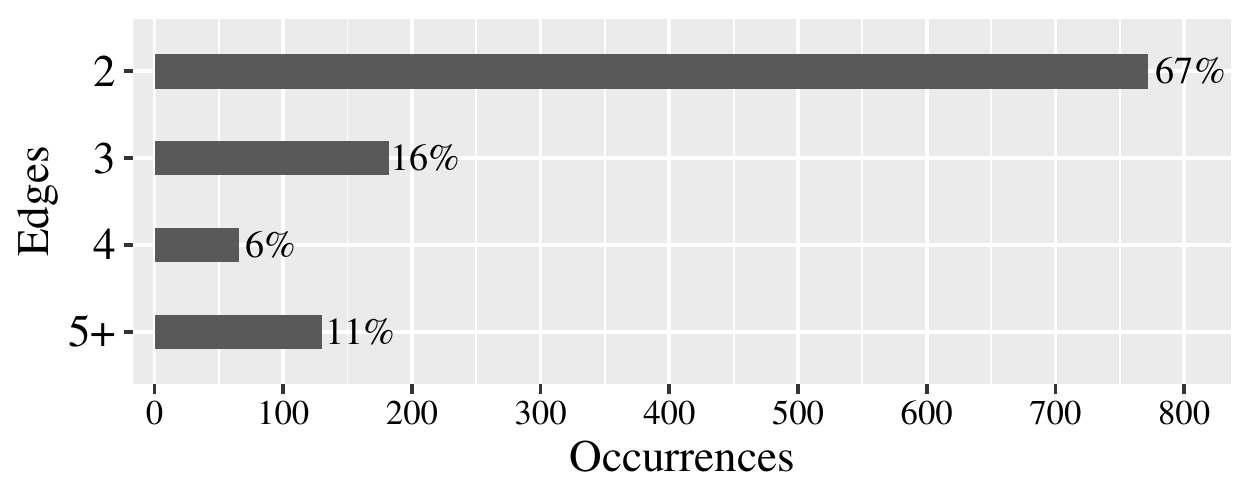}}
	\caption{Number of edges by refactoring subgraph}
	\label{fig:rq2-size-edges}
\end{figure}

As presented in Figure  \ref{fig:rq2-size-vertices}, most refactoring subgraphs have three vertices (639 occurrences, 56\%). 
The other recurrent cases comprise subgraphs with two (15\%) or four vertices (14\%).  
{\small \sffamily Square Okhttp} holds the largest subgraph regarding the number of vertices (57), which are most related to \textit{inline} operations.
Concerning the number of edges, most subgraphs have two (67\%) or three edges (16\%), as shown in Figure \ref{fig:rq2-size-edges}. 
{\small \sffamily MPAndroidChart} has the largest subgraph in term of edges.  
It has 61 edges, most representing \textit{extract and move} operations.
Therefore, most subgraphs contain few methods  (vertices) and refactoring operations (edges).

\begin{figure}[!ht]
	\centering
	%\vspace{0.2cm}
    \subfigure{\includegraphics[width=0.4\textwidth]{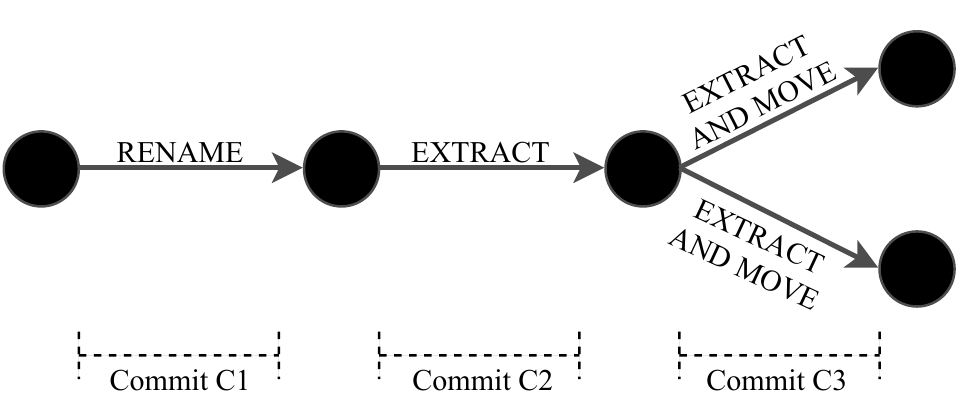}}
	\caption{Example of a refactoring subgraph from {\sffamily \scriptsize MPAndroidChart}}
	\label{fig:view_subgraph_PhilJay_MPAndroidChart_419}
\end{figure}

%_PhilJay_MPAndroidChart_419
Figure \ref{fig:view_subgraph_PhilJay_MPAndroidChart_419}  shows a real example of a refactoring subgraph from {\sffamily \small MPAndroidChart}, which includes three distinct refactoring operations.
In the first commit C1, a developer renamed method $drawYLegend()$ to $drawYLabels()$.\footnote{\url{https://github.com/PhilJay/MPAndroidChart/commit/13104b26}} 
In the subsequent operation performed 13 days later, the same developer extracted a new method from $drawYLabels()$ at commit C2.\footnote{\url{https://github.com/PhilJay/MPAndroidChart/commit/063c4bb0}}
Two days after the second operation, in commit C3, he made new extractions from $drawYLabels()$ to another class, creating a subgraph with five vertices and four edges.\footnote{\url{https://github.com/PhilJay/MPAndroidChart/commit/d930ac23}}\\[-0.2cm]

%elastic_elasticsearch_765
%elastic_elasticsearch_1280
%facebook_fresco_156

\begin{tcolorbox}[left=0mm,right=0mm,boxrule=0.25mm,colback=gray!5!white]
\vspace{-0.2cm}
{\em Summary:} Most refactoring subgraphs are small. Among 1,150 samples,  most cases comprise subgraphs with the number of vertices ranging from two to four  (85\%) and the number of edges varying between two and three (83\%).
%\vspace{-0.2cm}
\end{tcolorbox}

%-----------------------------------------
\subsection{(RQ2) How Many Commits Are in Refactoring Subgraphs?}
%tamanho

% --
% 3º Quartile
% facebook/fresco 			        75%       2.000000
% irbnb/lottie-android			    75%       2.000000
% spring-projects/spring-framework 	75%        2.000000
% square/retrofit			    	75%       2.000000
% ReactiveX/RxJava			        75%        2.000000

% bumptech/glide			    	75%        3.000000
% PhilJay/MPAndroidChart			75%       3.000000
% square/okhttp				        75%       3.000000
% elastic/elasticsearch			    75%        3.000000

% apache/dubbo				        75%       2.750000
% --------
% Valores maximos:
% elastic/elasticsearch 			max        9.000000
% square/okhttp				        max      18.000000

%square/okhttp; ID 238;18 commits - max
%spring-projects/spring-framework;ID 80; 9 commits - max

\begin{figure}[!ht]
	\centering
    \subfigure{\includegraphics[width=0.45\textwidth]{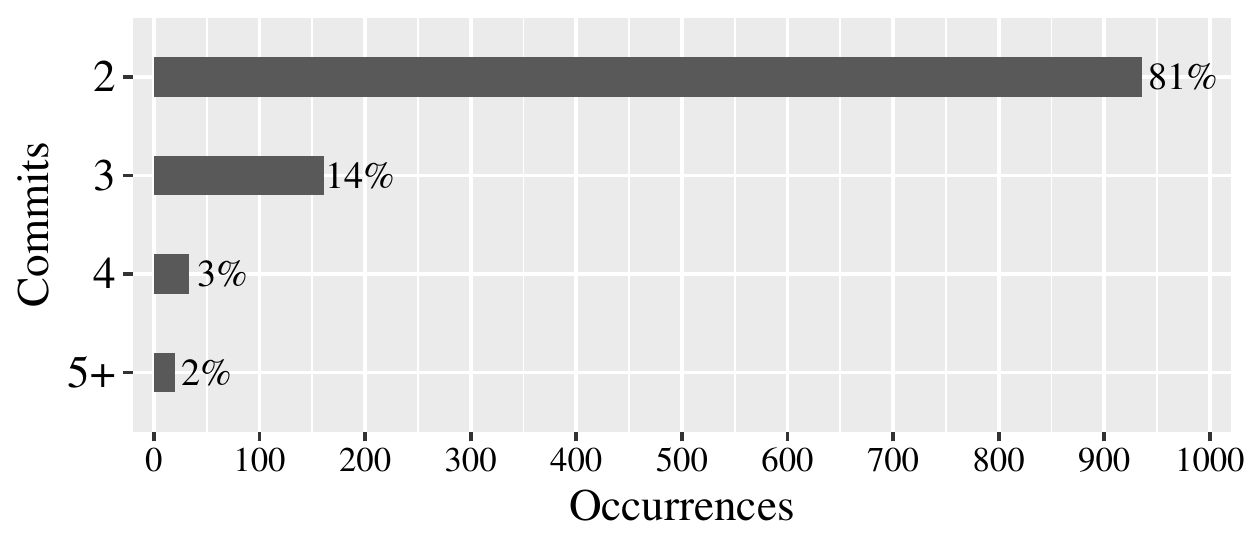}}
	\caption{Number of commits by refactoring subgraph}
	\label{fig:rq3-commits}
\end{figure}

In this second question, we investigate the number of commits per subgraph.  
%Our goal is to assess how often subgraphs increase over time. 
As presented in Figure~\ref{fig:rq3-commits}, most cases include subgraphs with two  (81\%)  or  three commits (14\%).
The largest subgraph in terms of commits is again from {\sffamily \small Square Okhttp} (18 commits).
%It is also the largest subgraph in terms of vertices, as stated in the previous RQ.

\begin{figure}[!ht]
	\centering
    \subfigure{\includegraphics[width=0.29\textwidth]{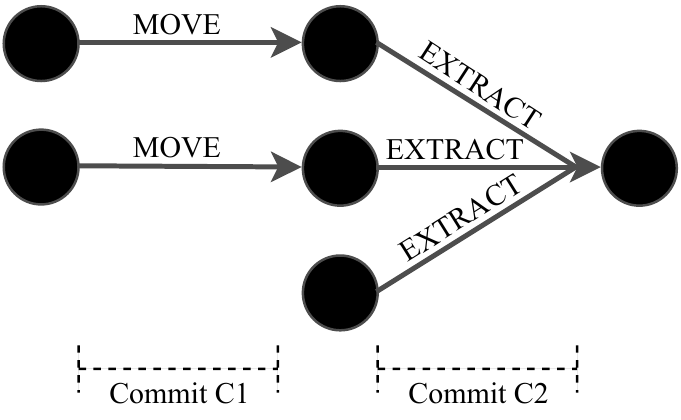}}
	\caption{Example of a refactoring subgraph from {\sffamily \scriptsize Elasticsearch}}
	\label{fig:view_subgraph_elastic_elasticsearch_767}
\end{figure}

Figure \ref{fig:view_subgraph_elastic_elasticsearch_767} shows an example from {\sffamily \small Elasticsearh}. 
In commit C1, a developer moved two methods from class $SocketSelector$ to $NioSelector$.\footnote{\url{https://github.com/elastic/elasticsearch/commit/9ee492a3f07}} 
After approximately three months, in commit C2, a second developer extracted duplicated code from three methods to a new method named $handleTask(Runnable)$.\footnote{\url{https://github.com/elastic/elasticsearch/commit/11fe52ad767}}  Among the source methods, two methods are the ones moved early. As a consequence, these two commits create a refactoring subgraph with six vertices and five edges.

%view_subgraph_elastic_elasticsearch_767

\begin{tcolorbox}[left=0mm,right=0mm,boxrule=0.25mm,colback=gray!5!white]
\vspace{-0.2cm}
{\em Summary:} Most refactoring subgraphs are created in two commits (81\%) or in three commits (14\%). 
\vspace{-0.2cm}
\end{tcolorbox}

%-----------------------------------------
\subsection{(RQ3) What Is the Age of Refactoring Subgraphs?}

\begin{figure}[!ht]
	\centering
    \subfigure{\includegraphics[width=0.49\textwidth]{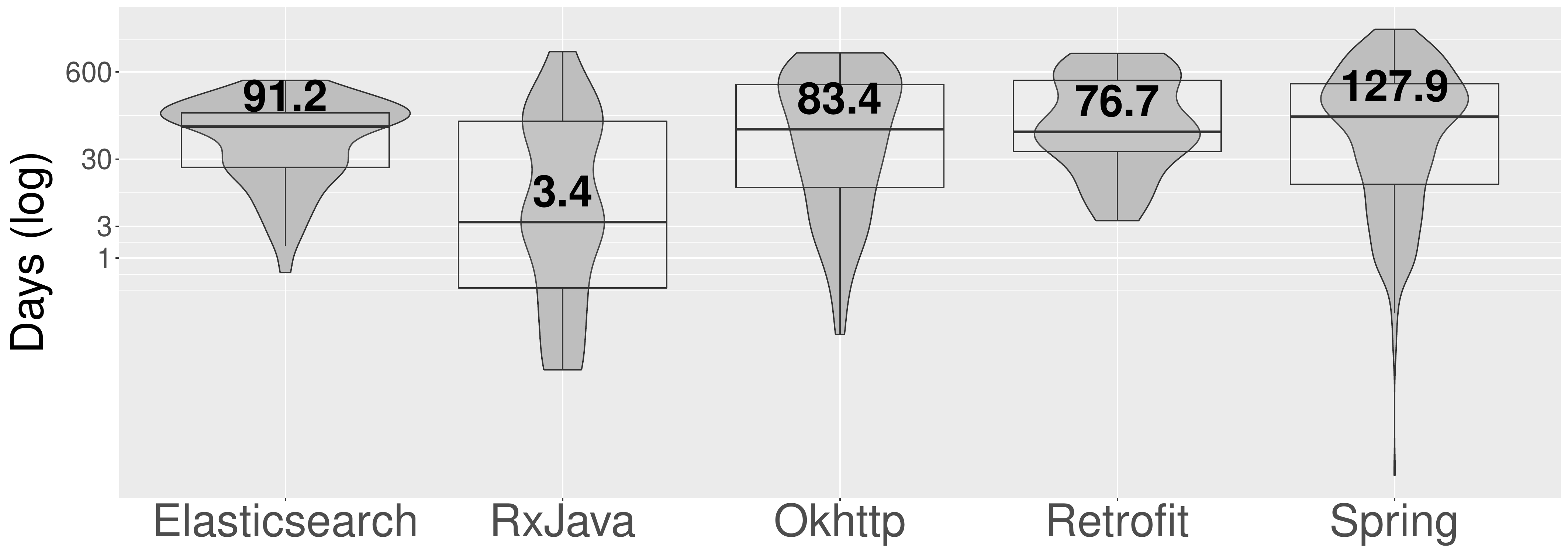}}
	\subfigure{\includegraphics[width=0.49\textwidth]{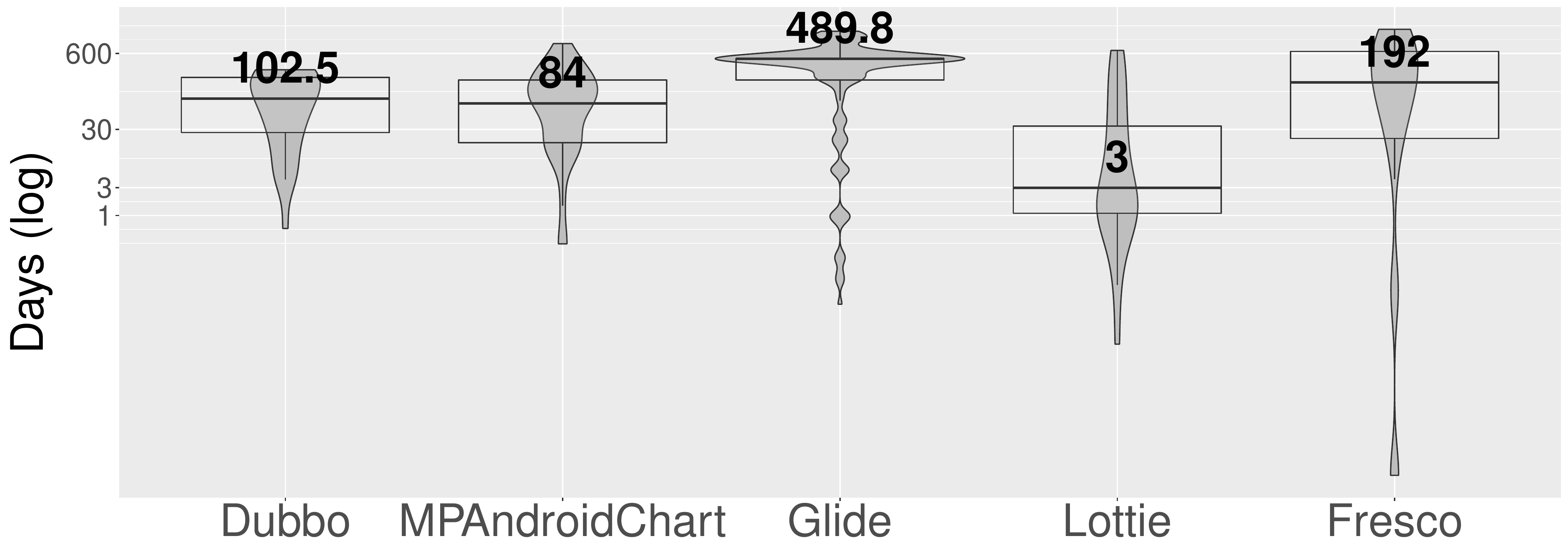}}
	\caption{Age of the refactoring subgraphs}
	\label{fig:rq4-age}
\end{figure}

To assess age, we compute the number of days between the most recent and the oldest commit in a refactoring subgraph.
Figure~\ref{fig:rq4-age} presents the results: we notice that refactoring subgraphs age varies among the projects.
Considering the median of the distributions, the youngest subgraphs are found in {\sffamily \small Lottie Android} and {\sffamily \small RxJava}, which have 3 and 3.4 days, respectively.
On the other side, the oldest subgraphs are found in {\sffamily \small Glide} (489.8 days), {\sffamily \small Spring Framework} (127.9), and Fresco (192).
The other systems have subgraphs with age between 76.7 ({\sffamily \small Retrofit}) and 102.5 days ({\sffamily \small Dubbo}).
Regarding the maturity of the target systems, the youngest project is {\sffamily \small Lottie Android} (3 years) while the oldest one is {\sffamily \small Elasticsearch} (9 years).
We run the Spearman's test to assess the correlation between the systems age and the median time of their refactoring subgraphs.
The correlation coefficient ($\mathrm{rho}$) is 0.067, showing a very weak correlation. % between system and subgraph age.
In other words, there are subgraphs with different age in both old and young systems.
However, the \textit{p-value} is $>$ 0.001 due to our small sample size.

%Sem remover outlier
%Escala de log
%Idade: |aresta mais nova - aresta mais antiga|

\begin{figure}[!ht]
	\centering
    \subfigure{\includegraphics[width=0.3\textwidth]{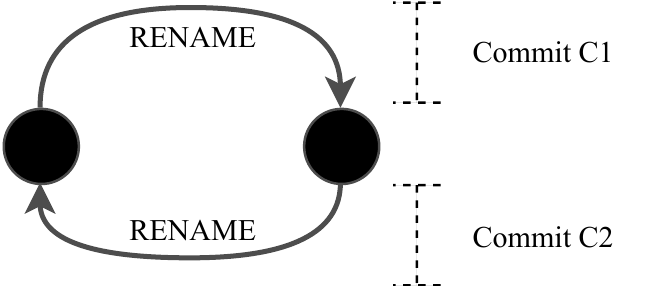}}
	\caption{Example of a refactoring subgraph from {\sffamily \scriptsize Spring Framework}}
	\label{fig:view_subgraph_spring_projects_spring_framework_375}
\end{figure}

Figure \ref{fig:view_subgraph_spring_projects_spring_framework_375} shows an example of a subgraph describing refactorings performed in few days on {\sffamily \small Spring Framework}. In commit C1, a  developer renamed method $before(Function)$ to $filterBefore(Function)$.\footnote{\url{https://github.com/spring-projects/spring-framework/commit/794693525f}}  After six days, the same developer reverted the operation in commit C2, renaming $filterBefore(Function)$ to the original name.\footnote{\url{https://github.com/spring-projects/spring-framework/commit/91e96d8084}} 
As a consequence, these modifications created a subgraph with two vertices and two edges.
%Indeed, the commit description suggests the team recommended to keep the original name: \textit{``This commit improves the RouterFunctions.Builder based on conversations had during the weekly team meeting''}.
%Overall, although we find some young subgraphs with few days, the majority has a lifetime of weeks or even months. 

%view_subgraph_spring_projects_spring_framework_375

\begin{tcolorbox}[left=0mm,right=0mm,boxrule=0.25mm,colback=gray!5!white]
\vspace{-0.2cm}
{\em Summary:} The age of the subgraphs is diverse: while some have few days, the majority of the subgraphs have weeks or even months. For example, 67\% of the refactoring subgraphs have more than one month.
\vspace{-0.2cm}
\end{tcolorbox}

%-----------------------------------------
\subsection{(RQ4) Which Refactorings Compose the Refactoring Subgraphs?}

%puro - subgrafo com arestas do mesmo tipo, mesma refatoração
%inpuro - subgrafo com arestas de tipos diferentes

First, we present the most common refactoring operations in our sample of 1,150 refactoring subgraphs (Table \ref{table:frequency_refactorings}).
Most cases include  \textit{rename method} (21\%), \textit{extract and move method} (19\%), and \textit{extract method} (17\%).
By constrast, we detected only 83 occurrences of \textit{move and rename} operations.
There are also few inheritance-based refactorings, i.e., \textit{pull up} (330 occurrences) and \textit{push down} (142 occurrences).

\begin{table}[!ht]
\centering
\caption{frequency of refactoring operations}
\label{table:frequency_refactorings}
\begin{tabular}{l r r}
\toprule
{\bf Refactoring} & {\bf Occurrences} & {\bf \%}
\\ \midrule

Rename	&	757	&	21	\\
Extract and move	&	685	&	19	\\
Extract	&	635	&	17	\\
Move	&	579	&	16	\\
Inline	&	474	&	13	\\
Pull up	&	330	&	9	\\
Push down	&	142	&	4	\\
Move and rename	&	83	&	2	\\

\midrule

All	&	3,685	&	100	\\

\bottomrule
\end{tabular}        
\end{table}

\begin{figure}[!ht]
    %\vspace{0.2cm}
	\centering
    \subfigure{\includegraphics[width=0.45\textwidth]{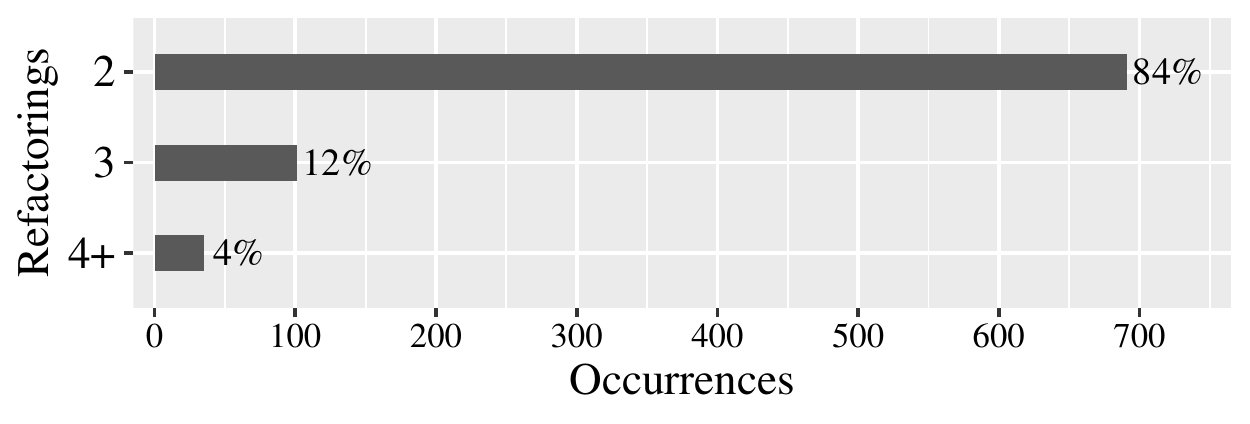}}
	\caption{Number of distinct refactoring operations in heterogeneous subgraphs}
	\label{fig:rq5_barplot_distinct_refactorings}
	\vspace{0.2cm}
\end{figure}

Next, we categorize the subgraphs into two groups. 
The homogeneous group includes subgraphs with a single refactoring operation.  In contrast, the heterogeneous category comprises subgraphs with at least two distinct refactoring operations. 
As presented in Table~\ref{table:rq5-composition}, overall, around 28\% of the subgraphs are homogeneous, while 72\% are heterogeneous.
%That is, the majority of the subgraphs are composed by at least two distinct refactoring operations.
The results per system follow a similar tendency. Most of the projects have more heterogeneous subgraphs than homogeneous ones; the sole exception is {\sffamily \small RxJava} (57\% vs 43\%).
%Homogeneity varies from 14.5\% to 57.1\%, while heterogeneity ranges from 42.9\% to 85.5\%.
In addition, as presented in Figure \ref{fig:rq5_barplot_distinct_refactorings}, heterogeneous subgraphs often include two distinct refactoring types (84\%); in contrast, 12\% have three and only 4\% have four or more distinct refactoring types.

\begin{table}[!ht]
%\vspace{-0.25cm}
\centering
\caption{homogeneous vs  heterogeneous refactoring subgraphs}
\label{table:rq5-composition}
\begin{tabular}{l r r r r }
\toprule
{\bf Project} & {\bf Homogeneous} & {\bf \%} & {\bf Heterogeneous} & {\bf \%}
\\ \midrule

Elasticsearch	&	43	&	30.9	&	96	&	69.1	\\
RxJava	&	56	&	57.1	&	42	&	42.9	\\
Square Okhttp	&	22	&	25.3	&	65	&	74.7	\\
Square Retrofit	&	12	&	35.3	&	22	&	64.7	\\
Spring Framework	&	138	&	27,6	&	362	&	72,4	\\
Apache Dubbo	&	6	&	20.0	&	24	&	80.0	\\
MPAndroidChart	&	16	&	21.9	&	57	&	78.1	\\
Glide	&	19	&	14.5	&	112	&	85.5	\\
Lottie Android	&	5	&	21.7	&	18	&	78.3	\\
Facebook Fresco	&	6	&	17.1	&	29	&	82.9	\\

\midrule

All	&	323	&	28.1	&	827	&	71.9	\\

\bottomrule
\end{tabular}        
\end{table}

\begin{figure}[!ht]
	\centering
	%\vspace{0.2cm}
    \subfigure{\includegraphics[width=0.40\textwidth]{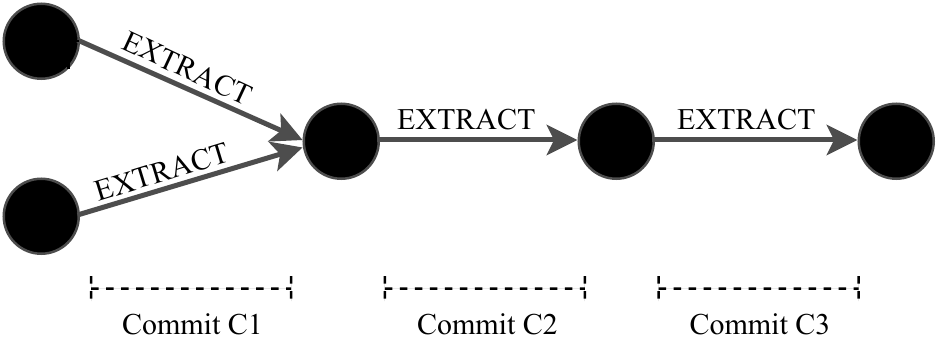}}
	\caption{Example of a homogeneous refactoring subgraph from {\sffamily \scriptsize Facebook Fresco}}
	\label{fig:view_subgraph_facebook_fresco_5}
\end{figure}

Figure \ref{fig:view_subgraph_facebook_fresco_5} shows an example of a homogeneous subgraph from {\sffamily \small Facebook Fresco}.
In this case, the subgraph represents four \textit{extract} operations performed over time.
First, in commit C1,  a developer extracted method $\mathrm{fetchDecodedImage(...)}$ from two methods into class $ImagePipeline$.\footnote{\url{https://github.com/facebook/fresco/commit/02ef6e0f}} 
The next operations happened years later when a second developer made two new \textit{extract} operations  in commits C2\footnote{\url{https://github.com/facebook/fresco/commit/b76f56ef}} and C3\footnote{\url{https://github.com/facebook/fresco/commit/017c007b}}.\\[-0.2cm]

%different_commits_view_subgraph_facebook_fresco_5.html
%C1 02ef6e0f6c9923b468ded065c824d68617eaec2a
%C2 b76f56ef976ba6b5d6176a2ea02c2261e3b1406d
%C3 017c007b48ed4eabb4c7e63be02bfc22e0a149ee

%As a second example, we presents a subgraph from \xxx, which includes \xxx distinct refactoring operations. First, the method \xxx...

\begin{tcolorbox}[left=0mm,right=0mm,boxrule=0.25mm,colback=gray!5!white]
\vspace{-0.2cm}
{\em Summary:} Most refactoring subgraphs are heterogeneous (71.9\%), i.e., they include more than one refactoring type. %In this category, most subgraphs have two distinct refactoring operations (84\%).
\vspace{-0.2cm}
\end{tcolorbox}

%-----------------------------------------
\subsection{(RQ5) Are the Refactoring Subgraphs Created by the Same or Multiple Developers?}

%mesmo desenvolvedor x desenvolvedores diferentes
%usamos o email do author para fazer a contagem.

As the last research question, we separate the refactoring subgraphs into two groups. 
The first group includes subgraphs with refactoring operations performed by a single developer. 
The second category is the opposite; it holds subgraphs by multiple developers.  
As presented in Table~\ref{table:rq6-dev}, most subgraphs have a single author (60.3\%).
%In fact, a recent study points out that overall few developers working in most issues on a system~\cite{Avelino:icpc:2016}.
As reported in a previous question,  the number of commits per subgraph is also small. 
Thus,  we execute Spearman's test to evaluate the correlation between the number of developers and the number of commits for each refactoring subgraph. The correlation coefficient ($\mathrm{rho}$) is 0.244, with a \textit{p-value} $<$ 0.001, indicating a weak correlation between these metrics.
That is, the higher the number of commits in a subgraph, the higher its amount of developers.

%https://ieeexplore.ieee.org/document/5645555
%https://ieeexplore.ieee.org/document/8449257/

\begin{table}[!ht]
%\vspace{-0.25cm}
\centering
\caption{Developers of refactoring graphs}
\label{table:rq6-dev}
\begin{tabular}{l r r r r}
\toprule
{\bf Project} & {\bf Single dev.} & {\bf \%} & {\bf Multiple devs.} & {\bf \%}
\\ \midrule

Elasticsearch	&	32	&	23.0	&	107	&	77.0	\\
RxJava	&	88	&	89.8	&	10	&	10.2	\\
Square Okhttp	&	32	&	36.8	&	55	&	63.2	\\
Square Retrofit	&	14	&	41.2	&	20	&	58.8	\\
Spring Framework	&	303	&	60.6	&	197	&	39.4	\\
Apache Dubbo	&	17	&	56.7	&	13	&	43.3	\\
MPAndroidChart	&	70	&	95.9	&	3	&	4.1	\\
Glide	&	116	&	88.5	&	15	&	11.5	\\
Lottie Android	&	11	&	47.8	&	12	&	52.2	\\
Facebook Fresco	&	10	&	28.6	&	25	&	71.4	\\

\midrule

All	&	693	&	60.3	&	457	&	39.7	\\

\bottomrule
\end{tabular}        
\end{table}

%different_commits_view_subgraph_square_okhttp_476.html
\begin{figure*}[!ht]
	\centering
    \subfigure{\includegraphics[width=0.95\textwidth]{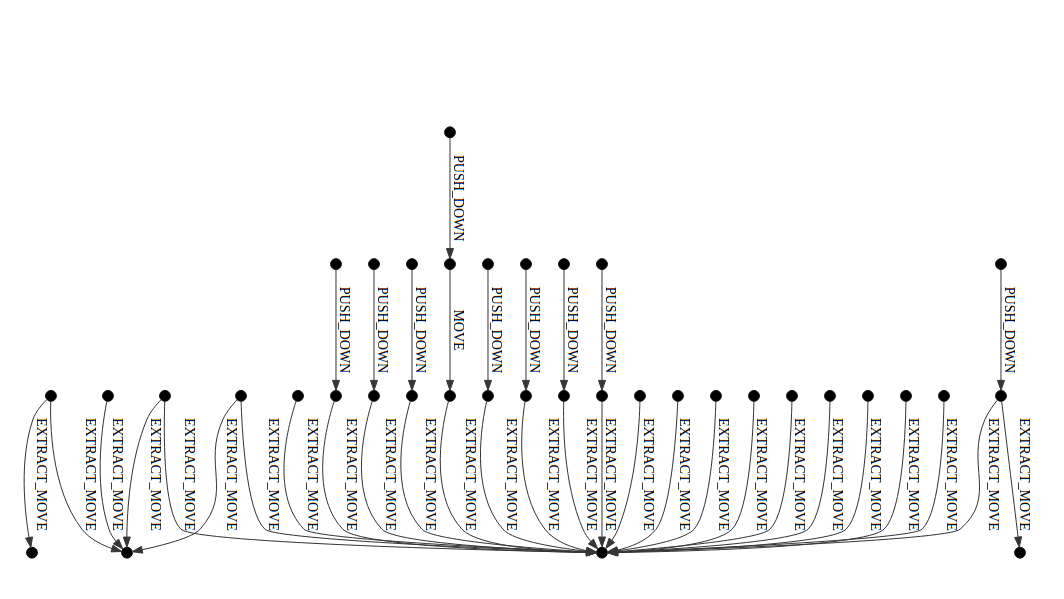}}
	\caption{Example of a large refactoring subgraph from {\sffamily \scriptsize Square Okhttp}}
	\label{fig:view_subgraph_square_okhttp_476}
\end{figure*}

\begin{figure}[!ht]
	\centering
    \subfigure{\includegraphics[width=0.4\textwidth]{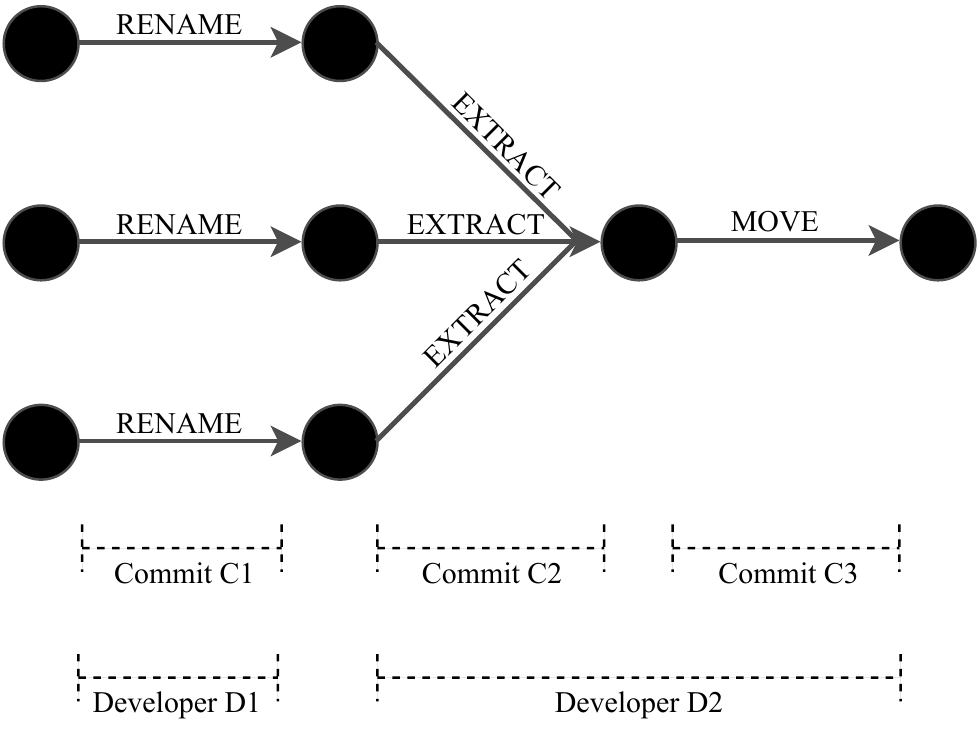}}
	\caption{Example of a refactoring subgraph create by multiple developers from {\sffamily \scriptsize Square Okhttp}}
	\label{fig:view_subgraph_square_okhttp_85}
\end{figure}

Figure \ref{fig:view_subgraph_square_okhttp_85} presents an example of a refactoring subgraph from {\sffamily \small Square Okhttp}. 
First, in commit C1, a developer D1 renamed three methods from class $OkHttpClient$.\footnote{\url{https://github.com/square/okhttp/commit/daf2ec6b9}} 
Basically, the developer removed the prefix $set$ from their names.
%$setConnectTimeout(...)$ to $connectTimeout(...)$, $setWriteTimeout(...)$ to $writeTimeout(...)$, and $setWriteTimeout(...)$ to $writeTimeout(...)$.   
After 10 months, a second developer D2 removed a duplicate code from these methods, extracting method $checkDuration(...)$.\footnote{\url{https://github.com/square/okhttp/commit/c5a26fefd}}  
Then, after seven months, D2 moved this method to a new class named $Util$, in commit C3.\footnote{\url{https://github.com/square/okhttp/commit/a32b1044a}} As a result, these two developers were responsible for a refactoring subgraph with eight vertices and seven edges.  \\[-0.2cm]

%view_subgraph_square_okhttp_85

\begin{tcolorbox}[left=0mm,right=0mm,boxrule=0.25mm,colback=gray!5!white]
\vspace{-0.2cm}
{\em Summary:} Most refactoring subgraphs are created by a single developer (60\%). Only 40\% have multiple developers.
\vspace{-0.2cm}
\end{tcolorbox}

%\todo{teste Spearman}
%https://geographyfieldwork.com/SpearmansRankCalculator.html
%TODO: Incluir texto sobre o teste estatístico.
%Resultado do teste estatístico:
%Para todos os subgrafos, entrada com 1,150 linhas/subgrafos.
% número de commits distintos x número de desenvolvedores
% conf.level = 0.99
% Spearman's rank correlation rho
% data:  dataset$count_commits and dataset$count_developers
% S = 191700000, p-value < 2.2e-16
% alternative hypothesis: true rho is not equal to 0
% sample estimates:
%       rho 
% 0.2437136 

\section{Large Subgraph Example}
\label{sec:example}

In this section, we present and discuss an example of a large refactoring subgraph. As we reported in Section~\ref{section:results}, most refactoring subgraphs are small, in terms of number of vertices, edges, and commits. For this reason, we only presented small examples when discussing our RQ results. However, we also found graphs describing major refactorings over time, whose presentation we postponed to this section.

Figure~\ref{fig:view_subgraph_square_okhttp_476} shows an example from {\sffamily \small Square Okhttp}. 
We chose this example because it encompasses different refactoring operations performed over time and it is one of the largest subgraphs from our dataset.
This graph has 37 vertices, four commits, and three refactoring operations (\textit{move}, \textit{push down}, and \textit{extract and move}). 
It was built by multiple developers, over six months. %183 days
As we can observe, the graph nicely describes an example of code duplication removal.
First, a developer performed nine {\em push down} refactorings to move a method from a superclass to a subclass. 
Then, a second developer performed 21 {\em extract method} operations to move the duplicated code to a single method, which has the following code:

\begin{small}
\begin{verbatim}
public int readInt() throws IOException {
  require(4, Deadline.NONE);
  return buffer.readInt();
}
\end{verbatim}
\end{small}

Besides that, there are other three {\em extract method} operations: (i) {\em readShort()} from a single
method (this node has a single incoming edge), (ii) {\em readByteString()} from four methods, and (iii) {\em decode()} from a single method.
These new methods are presented in the bottom of Figure~\ref{fig:view_subgraph_square_okhttp_476}.

%-----------------------------------------
\section{Discussion and Implications}
\label{section:discussion}

\subsection{Detecting Refactoring over Time}

Several tools and techniques are proposed in the literature to detect refactoring operations, for instance, Refactoring Crawler~\cite{Dig:2006:ECOOP}, RefFinder~\cite{Kim:FSE:2010}, Refactoring Miner~\cite{Tsantalis:2013:CASCON,danilo:fse2016:WhyWeRefactor}, and, more recently, RefDiff~\cite{danilo:msr2017:RefDiff} and RMiner~\cite{Tsantalis:ICSE:2018:RefactoringMiner}.
In common, those approaches only detect \emph{atomic} refactoring, i.e., operations that happen in a single commit and performed by a single developer.
In contrast, our approach, refactoring graph, focuses on the detection of refactoring over time, i.e., operations over multiple commits and performed by multiple developers.
Moreover, differently from the \emph{batch} refactoring~\cite{Murphy-Hill:ICSE:2009, Bibiano:esem2019:BatchRefactoring, cedrimRego:2018}, our approach is not constrained by the amount of developers nor to a time window.
Indeed, we found refactoring subgraphs with age ranging from weeks to months and created by multiple developers.
\emph{Therefore, we contribute to the refactoring literature with a novel approach to detect and explore refactoring operations in a broader perspective to complement existing tools and techniques.}

\subsection{Refactoring Comprehension and Improvement}

When performing code review, developers often adopt diff tools to better understand code changes, and decide whether they will be accepted or not.
In this process, developers may also look for defects and code improvement opportunities~\cite{bacchelli2013expectations}.
However, if the reviewed change is large and complex, this task becomes challenging~\cite{bacchelli2013expectations}.
To alleviate this issue, refactoring-aware code review tools were proposed~\cite{hayashi2013rediffs, ge2014towards, ge2017refactoring} to better understand changes mixed with refactoring.
Refactoring graphs can contribute to handle this issue by providing navigability at method level. 
That is, a code reviewer may navigate back in a method to reason how a similar change was performed.
For example, in Figure~\ref{fig:view_subgraph_square_okhttp_85}, a code reviewer may investigate whether all methods were properly renamed in the past, before accepting commit C3.
\emph{Thus, refactoring graphs can be integrated to code review tools to better support code understating and improvement.}

\subsection{Detecting Refactoring Patterns and Smells}

Frequent refactoring subgraphs may indicate common refactoring patterns over time.
In contrast, infrequent refactoring subgraphs that are variations of the pattern may suggest the presence of ``refactoring smells'' that deserve to be fixed.
For example, suppose the refactoring subgraph shown in Figure~\ref{fig:example_refactoring_graph_3_complexo} is frequent:
a developer extracted two methods from $m2()$, which are named $a()$ and $b()$;
then, $b()$ was renamed to $c()$, finally, $c()$ was renamed back to $b()$.
In this case, if we find a single refactoring subgraph that does not include the last renaming, this may suggest that the developer forgot to perform the undo rename in one single case.
In this sense, refactoring subgraphs can be used to spot bad smells, which are only visible because refactoring subgraphs provide the big picture of the refactoring.
Indeed, this is a topic that we aim to deep assess in further research, possibly with the support of techniques to mine graphs~\cite{Yan:2002:gspan,Inokuchi:2019:AGM,Kuramochi:2001:FSG}.
\emph{Thus, refactoring graphs can foment the detection of refactoring anomalies over time and drive future research agenda on refactoring patterns.}

\subsection{Understanding and Assessing Software Evolution}

During software evolution, developers often perform refactoring operations.
Consequently, the link between methods may be lost~\cite{hora:icse2018:UntrackedChanges}.
For example, if a method $a()$ is renamed to $b()$ and then extracted to $c()$, it becomes quite hard to trace $a()$ to $c()$, and vice versa.
This has several implications to software evolution research, particularly on studies that assess multiple code versions, such as code authorship detection~\cite{Avelino:icpc:2016, rahman11, meneely12,spinellis17, hattori09}, code evolution visual supporting~\cite{gomez10, gomez15}, bug introducing change detection~\cite{Kim06a, Zimm06a, rahman11a, chen2014empirical, ray16}, to name a few.
In practice, these studies often rely on tools provided by Git and SVN, such as \texttt{git blame} and \texttt{svn blame}, which show what revision and author last modified each line of a file.
However, this process is sensitive to refactoring operations~\cite{Avelino:icpc:2016, hora:icse2018:UntrackedChanges}.
As Git and SVN tools cannot track fine-grained refactoring operations, particularly at method level, these approaches may miss relevant data.
For instance, in the aforementioned example, it would be not possible to detect that method $c()$ was originated in method $a()$. Consequently, we would be not able to find the real creator of method $c()$ nor the developer who introduced a bug on $c()$.
\emph{With refactoring graphs, we are able to resolve method names over time, thus, software evolution studies can benefit as more precise tools can be created on the top.}

\section{Threats to Validity}
\label{section:threatsValidity}

\noindent\emph{Generalization of the results.}
We analyzed 1,150 refactoring subgraphs from 10 popular and open source Java systems.
Therefore, our dataset is built over credible and real-world software systems.
Despite these observations, our findings---as usual in empirical software engineering---may not be directly generalized to other systems, particularly commercial, closed source, and the ones implemented in other languages than Java.
Besides that, we focus our study on eight refactorings at method level. Thus, other refactoring types can affect the size of subgraphs.
We plan to extend this research to cover software systems implemented in other programming languages and refactorings at class level.\\[-0.2cm]

\noindent\textit{Adoption of \refdiff.}
We adopted \refdiff~to detect refactoring operation because it is the sole refactoring detection tool that is multi language, working for Java, JavaScript, and C.
It is also extensible to other programming languages.
Thus, as we plan to extend this research to cover other programming languages than Java, \refdiff~was the proper solution.
In addition to be multi language, \refdiff~accuracy is quite high.
\refdiff's authors provide two evaluations of their tool~\cite{danilo:msr2017:RefDiff}.
In the first evaluation, it achieved an overall F-measure of 96.8\% (precision: 100\%; recall: 93.9\%).
In the second evaluation, \refdiff's authors analyzed 102 real refactoring instances.
In this case, it achieved an overall F-measure of 89.3\% (precision: 85.4\%; recall: 93.6\%).
Recently, Tsantalis~\textit{et al.}~\cite{Tsantalis:ICSE:2018:RefactoringMiner} proposed the refactoring detection tool {\sc rminer}.
When considering all refactoring operations, {\sc rminer} has an F-measure of 92\% (precision: 98\%; recall: 87\%) improving on \refdiff's overall accuracy.
However, {\sc rminer} works only for Java projects.\\[-0.2cm]

\noindent\textit{Building refactoring graphs.}
When creating the refactoring graphs, we cleaned up our data (i.e., vertices and edges) to keep only meaningful subgraphs.
For instance, we removed constructor methods (vertices) from our analysis because they include mostly initialization settings, and do not have behavior as conventional methods.
We also removed some very specific cases of refactoring (edges) in which \refdiff~reported false positives in inner classes or same method.
However, these cases are not likely to affect our results because they only represent a fraction (3.5\%) of the refactoring operations.
Finally, the refactoring subgraphs can include unintentional operations (e.g., reverted commits by automatic deployment systems). To mitigate this threat, we focus our study on the main branch evolution to avoid experimental or unstable versions.\\[-0.2cm]
%However, these cases are not likely to affect our results because they only represent a fraction (3.5\%) of the refactoring operations.

%(382+102)÷(13756) = 3.5%
%Total edges: 13756
%Removed equals nodes (m1 and m2): 102
%Removed constructors (edges): 514
%Removed duplicated edges: 382

\noindent\textit{Detection of developers.}
In RQ5, we investigate the number of developers per refactoring subgraphs. 
We used the email available on {\sffamily \small git log} to distinguish the developers. 
Thus, our results can include the same developer committing with different email addresses. 
But, we already found that most cases are subgraphs created by a single developer. 
%In fact, a recent study points out that, overall, few developers work on most issues on a system~\cite{Avelino:icpc:2016}.
%Finally, since we analyze the main branch,  we did not retrieve the author of refactoring operations performed in another branches.
%}

%-----------------------------------------
\section{Related Work}
\label{section:relatedWork}

Refactoring is an usual practice during software evolution and maintenance.
Constantly, developers  refactor the source code for different purposes~\cite{danilo:fse2016:WhyWeRefactor,Wang:ICSM:2019}.
For this reason, several studies concentrate on this research field~\cite{Murphy-Hill:ICSE:2009,Bibiano:esem2019:BatchRefactoring,Mahmoudi:2019:SANER,  Lin:SANER:2019,Tsantalis:ICSE:2018:RefactoringMiner,Dig:2006:ECOOP, danilo:msr2017:RefDiff,Kim:2014:TSE,Kim:icse:2016,Szoke:SANER:2016,Bavota:JournalSystem:2015,Bavota:SCAM:2012,johnson:acm2006,Shen:OOPSLA:2019,jss-2018-jmove,Alves:2014:FSE,Lin:2016:FSE,Chaparro:ICSME:2014}. 
Among the empirical studies, some research focus on set of related refactoring.
Specifically, these studies analyze \textit{batch refactorings}~\cite{Murphy-Hill:ICSE:2009,Bibiano:esem2019:BatchRefactoring, Fernandes:2019:ICSE, Tenorio:2019:IWOR, Fernandes:2019:IWOR, cedrimRego:2018}. 
Murphy \textit{et al.}~\cite{Murphy-Hill:ICSE:2009} analyzed four datasets from different sources, all of these including metadata about the usage of Eclipse IDE.  
For instance, the dataset named {\em Everyone} contains Eclipse refactoring commands used by developers.   
Based on these datasets, the authors discuss usage and configurations of refactoring tools,  frequency of refactoring operations,  and commit messages. 
They also investigate about sets of refactorings operations executed in 60 seconds of each another, which are named \textit{batches}.  
The authors state that the some refactorings types are more common in batches, such as \textit{rename}, \textit{introduce a parameter}, and \textit{encapsulate field}.  
Besides that, about 47\% of refactorings performed using a refactoring tool happen in batches.  
However,  the baches involve a short period,  the study does not investigate refactorings operations that occur in different moments over time.

In another context, Bibiano \textit{et al.}~\cite{Bibiano:esem2019:BatchRefactoring} point  out that sets of related refactorings can solve problems due to code smells.  
The authors studied 54 GitHub projects and three closed systems. 
First, they used \textit{RMiner} tool to detect 13 well-know refactorings~\cite{Tsantalis:ICSE:2018:RefactoringMiner}, resulting in 24,893 operations. Then, the authors applied a heuristic to compute batch refactorings, i.e., set of related refactorings~\cite{cedrimRego:2018}. 
The heuristic includes two main requirements do retrieve a batch refactoring: (i) there are more than two refactoring operations in a single entity and (ii) the operations are from a single developer.  
The results are 4.607 batch refactorings. 
Next, the authors used another tool and scripts to identify more than 41K code smell occurrences in these systems. 
Finally, the authors computed the effect of batch refactorings to remove code smells. 
The main results show that most batches have only one commit (93\%) and two refactoring types. Also, the authors state that batches have a negative or neutral effect on code smells (81\%).
However, the authors focus on code smells and operations performed by a single developer.
In our study, the subgraphs involve refactoring over time (i.e., more than one commit), including subgraphs by multiples developers and different code elements. 

Other studies also discuss the impact of batches to eliminate code smells, proposing approaches to reuse or suggest sets of related refactoring operations~\cite{Tenorio:2019:IWOR, Fernandes:2019:IWOR,Jiau:TSE:2013}. 
Thus, they do not focus on sequences of refactoring operations over time. 
Fowler~\cite{Fowler:1999} mention a similar term called \textit{big refactoring}. 
The author points out that some refactorings are atomic, i.e.,  they are finished in a few minutes. 
By contrast, there are big refactorings, which are performed during months or years. 
We reinforce this observation: the age of the refactoring subgraphs is diverse, ranging from days to weeks or even months.

Hora~\textit{et al.}~\cite{hora:icse2018:UntrackedChanges} analyze untracked changes during software development. The authors show that refactorings invalidate several tracking strategies to evaluate system evolution. 
As in our study, they represent evolutionary changes as graphs. In this case, each node refers to a class or a method, and the edges indicate tracked changes (i.e., entities that keep their names after a modification) and untracked changes (i.e., entities that change their names after a refactoring). 
In other words, a graph represents traceable changes or alterations that split the entity's history. 
The results point up to 21\% of the changes at the method level and up to 15\% at the class level are untraceable. 
By contrast, in our study, the goal is to investigate refactorings performed over long time windows; we do not concentrate on tracked modifications on source code.

Meananeatra~\cite{Meananeatra:ASE:2012} also reports changes during software evolution as graphs. 
However, the study concentrates on refactoring sequences to remove {\em long methods}. 
The author proposes an approach based on two main criteria to detect an optimal set of refactorings.
An optimal refactoring sequence centers on four metrics: number of removed bad smells,  size of the refactoring sequence, number of the affected code elements, and the maintainability value (i.e., analyzability, changeability, stability, and testability).
The technique represents candidate refactoring sequences as graphs.  
In this case,  a graph contains a root node representing the original method version with smells.  
Each new node denotes a new method version after a refactoring operation. 
As in our study, the edges refer to refactorings.
By contrast, the nodes represent the same method before and after the changes. 
Each path in the graph is a candidate refactoring sequence, which can meet the selection criteria. 
Thus, the study does not focus on real refactorings over time.
Instead, the graph model represents steps to decompose a long method.

%-----------------------------------------
\section{Conclusion}
\label{section:conclusion}

In this paper, we proposed refactoring graphs, a novel approach to assess refactoring operations over time.
We analyzed 10 popular Java systems from which 1,150 refactoring  subgraphs were extracted.
We then investigate five research questions to evaluate the following properties of refactoring graphs: size, commits, age, composition, and developers.
We summarize our findings as follows:\\[-0.2cm]

\begin{itemize}
    \item The majority of the refactoring subgraphs are small (four nodes and three edges). However, there also outliers with dozens of nodes and edges.
    \item Most refactoring subgraphs have up to three commits (95\%).
    \item Refactoring subgraphs span from few days to months.
    \item Refactoring graphs are often heterogeneous, that is, they are composed by several types of refactoring. 
    \item Refactoring graphs are mostly created by a single developer (60\%).
\end{itemize}

Based on our findings, we provided further discussion and implications to our study.
Particularly,
(i) we discuss our contributions regarding refactoring tools as a novel approach to explore refactoring operations in a broader perspective;
(ii) we argue that refactoring graphs can be integrated to code review tools to better support code comprehension;
(iii) we claim that refactoring graphs can play a role on the detection of refactoring patterns and anomalies, only possible to be spotted over time; and
(iv) we state the importance of refactoring graphs to resolve method names and support software evolution studies.

Further studies can consider other popular programming languages and ecosystems; refactoring graphs based on class and package level as well as other refactoring types at method level; and also novel approaches to complement existing tools and techniques that focus on \textit{atomic} refactorings. \\[-0.2cm]

\section*{Acknowledgments}

%\noindent This research is supported by grants from [\textit{omitted due to double-blind review}].

\noindent This research is supported by grants from FAPEMIG, CNPq, and CAPES.

\balance

\bibliographystyle{ieeetr}
\bibliography{bib}

\begin{thebibliography}{10}

\bibitem{Murphy-Hill:ICSE:2009}
E.~{Murphy-Hill}, C.~Parnin, and A.~P. Black, ``How we refactor, and how we
  know it,'' in {\em 31st International Conference on Software Engineering
  (ICSE)}, pp.~287--297, 2009.

\bibitem{Negara:2013:ECOOP}
S.~Negara, N.~Chen, M.~Vakilian, R.~E. Johnson, and D.~Dig, ``A comparative
  study of manual and automated refactorings,'' in {\em 27th European
  Conference on Object-Oriented Programming (ECOOP)}, pp.~552--576, 2013.

\bibitem{danilo:fse2016:WhyWeRefactor}
D.~Silva, N.~Tsantalis, and M.~T. Valente, ``Why we refactor? {Confessions} of
  {GitHub} contributors,'' in {\em 24th International Symposium on the
  Foundations of Software Engineering (FSE)}, pp.~858--870, 2016.

\bibitem{Mazinanian:2017:OOPSLA}
D.~Mazinanian, A.~Ketkar, N.~Tsantalis, and D.~Dig, ``Understanding the use of
  lambda expressions in {Java},'' {\em Programming Languages}, vol.~1, no.~85,
  pp.~85:1--85:31, 2017.

\bibitem{Tsantalis:2013:CASCON}
N.~Tsantalis, V.~Guana, E.~Stroulia, and A.~Hindle, ``A multidimensional
  empirical study on refactoring activity,'' in {\em 23th Conference of the
  Center for Advanced Studies on Collaborative Research (CASCON)},
  pp.~132--146, 2013.

\bibitem{Kim:2012:FSE}
M.~Kim, T.~Zimmermann, and N.~Nagappan, ``A field study of refactoring
  challenges and benefits,'' in {\em 20th International Symposium on the
  Foundations of Software Engineering (FSE)}, pp.~50:1--50:11, 2012.

\bibitem{Kim:2014:TSE}
M.~Kim, T.~Zimmermann, and N.~Nagappan, ``An empirical study of refactoring
  challenge and benefits at {Microsoft},'' {\em Transactions on Software
  Engineering}, vol.~40, no.~7, pp.~633--649, 2014.

\bibitem{Bibiano:esem2019:BatchRefactoring}
A.~C. Bibiano, E.~F. D. O.~A. Garcia, M.~Kalinowski, B.~Fonseca, R.~Oliveira,
  A.~Oliveira, and D.~Cedrim, ``A quantitative study on characteristics and
  effect of batch refactoring on code smells,'' in {\em 13th International
  Symposium on Empirical Software Engineering and Measurement (ESEM)},
  pp.~1--11, 2019.

\bibitem{cedrimRego:2018}
D.~Cedrim, {\em Understanding and improving batch refactoring in software
  systems}.
\newblock PhD thesis, PUC-Rio, 2018.

\bibitem{Fowler:1999}
M.~Fowler, {\em Refactoring: Improving the Design of Existing Code}.
\newblock Addison-Wesley, 1999.

\bibitem{hudson:icsme2016:Popularity}
H.~Borges, A.~Hora, and M.~T. Valente, ``Understanding the factors that impact
  the popularity of {GitHub} repositories,'' in {\em 32nd International
  Conference on Software Maintenance and Evolution (ICSME)}, pp.~334--344,
  2016.

\bibitem{jss-2018-github-stars}
H.~Silva and M.~T. Valente, ``What's in a {GitHub} star? {Understanding}
  repository starring practices in a social coding platform,'' {\em Journal of
  Systems and Software}, vol.~146, pp.~112--129, 2018.

\bibitem{danilo:msr2017:RefDiff}
D.~Silva and M.~T. Valente, ``{RefDiff}: Detecting refactorings in version
  histories,'' in {\em 14th International Conference on Mining Software
  Repositories (MSR)}, pp.~1--11, 2017.

\bibitem{Dig:2006:ECOOP}
D.~Dig, C.~Comertoglu, D.~Marinov, and R.~Johnson, ``Automated detection of
  refactorings in evolving components,'' in {\em 20th European Conference on
  Object-Oriented Programming (ECOOP)}, pp.~404--428, 2006.

\bibitem{Kim:FSE:2010}
M.~Kim, M.~Gee, A.~Loh, and N.~Rachatasumrit, ``Ref-finder: a refactoring
  reconstruction tool based on logic query templates,'' in {\em 8th
  International Symposium on Foundations of software engineering (FSE)},
  pp.~371--372, 2010.

\bibitem{Tsantalis:ICSE:2018:RefactoringMiner}
N.~Tsantalis, M.~Mansouri, L.~M. Eshkevari, D.~Mazinanian, and D.~Dig,
  ``Accurate and efficient refactoring detection in commit history,'' in {\em
  40th International Conference on Software Engineering (ICSE)}, pp.~483--494,
  2018.

\bibitem{bacchelli2013expectations}
A.~Bacchelli and C.~Bird, ``Expectations, outcomes, and challenges of modern
  code review,'' in {\em 35th International Conference on Software Engineering
  (ICSE)}, pp.~712--721, 2013.

\bibitem{hayashi2013rediffs}
S.~Hayashi, S.~Thangthumachit, and M.~Saeki, ``{Rediffs}: Refactoring-aware
  difference viewer for {Java},'' in {\em 20th Working Conference on Reverse
  Engineering (WCRE)}, pp.~487--488, 2013.

\bibitem{ge2014towards}
X.~Ge, S.~Sarkar, and E.~Murphy-Hill, ``Towards refactoring-aware code
  review,'' in {\em 7th International Workshop on Cooperative and Human Aspects
  of Software Engineering (CHASE)}, pp.~99--102, ACM, 2014.

\bibitem{ge2017refactoring}
X.~Ge, S.~Sarkar, J.~Witschey, and E.~Murphy-Hill, ``Refactoring-aware code
  review,'' in {\em Symposium on Visual Languages and Human-Centric Computing
  (VL/HCC)}, pp.~71--79, 2017.

\bibitem{Yan:2002:gspan}
{Xifeng Yan} and {Jiawei Han}, ``{gSpan}: graph-based substructure pattern
  mining,'' in {\em 2nd International Conference on Data Mining (ICDM)},
  pp.~721--724, 2002.

\bibitem{Inokuchi:2019:AGM}
A.~Inokuchi, T.~Washio, and H.~Motoda, ``An apriori-based algorithm for mining
  frequent substructures from graph data,'' in {\em 4th Principles and Practice
  of Knowledge Discovery in Databases (PKDD)}, pp.~13--23, 2000.

\bibitem{Kuramochi:2001:FSG}
M.~Kuramochi and G.~Karypis, ``Frequent subgraph discovery,'' in {\em 1st
  International Conference on Data Mining (ICDM)}, pp.~313--320, 2001.

\bibitem{hora:icse2018:UntrackedChanges}
A.~Hora, D.~Silva, R.~Robbes, and M.~T. Valente, ``Assessing the threat of
  untracked changes in software evolution,'' in {\em 40th International
  Conference on Software Engineering (ICSE)}, pp.~1102--1113, 2018.

\bibitem{Avelino:icpc:2016}
G.~Avelino, L.~Passos, A.~Hora, and M.~T. Valente, ``A novel approach for
  estimating truck factors,'' in {\em 24th International Conference on Program
  Comprehension (ICPC)}, pp.~1--10, 2016.

\bibitem{rahman11}
F.~Rahman and P.~Devanbu, ``Ownership, experience and defects: a fine-grained
  study of authorship,'' in {\em 33rd International Conference on Software
  Engineering (ICSE)}, 2011.

\bibitem{meneely12}
A.~Meneely and O.~Williams, ``Interactive churn metrics: socio-technical
  variants of code churn,'' {\em Software Engineering Notes}, vol.~37, no.~6,
  2012.

\bibitem{spinellis17}
D.~Spinellis, ``A repository of {Unix} history and evolution,'' {\em Empirical
  Software Engineering}, vol.~22, no.~3, pp.~1372--1404, 2017.

\bibitem{hattori09}
L.~Hattori and M.~Lanza, ``Mining the history of synchronous changes to refine
  code ownership,'' in {\em 6th International Working Conference on Mining
  Software Repositories (MSR)}, 2009.

\bibitem{gomez10}
V.~U. G{\'o}mez, S.~Ducasse, and T.~D'Hondt, ``Visually supporting source code
  changes integration: the {Torch} dashboard,'' in {\em 17th Working Conference
  on Reverse Engineering (WCRE)}, 2010.

\bibitem{gomez15}
V.~U. G{\'o}mez, S.~Ducasse, and T.~D'Hondt, ``Visually characterizing source
  code changes,'' {\em Science of Computer Programming}, vol.~98, no.~P3,
  pp.~376--393, 2015.

\bibitem{Kim06a}
S.~Kim, T.~Zimmermann, K.~Pan, and E.~J.~J. Whitehead, ``Automatic
  identification of bug-introducing changes,'' in {\em 21st International
  Conference on Automated Software Engineering (ASE)}, 2006.

\bibitem{Zimm06a}
T.~Zimmermann, S.~Kim, A.~Zeller, and E.~J. Whitehead, Jr., ``Mining version
  archives for co-changed lines,'' in {\em 3rd International Workshop on Mining
  Software Repositories (MSR)}, 2006.

\bibitem{rahman11a}
F.~Rahman, D.~Posnett, A.~Hindle, E.~Barr, and P.~Devanbu, ``{BugCache} for
  inspections: hit or miss?,'' in {\em 19th International Symposium on the
  Foundations of Software Engineering (FSE)}, 2011.

\bibitem{chen2014empirical}
T.-H. Chen, M.~Nagappan, E.~Shihab, and A.~E. Hassan, ``An empirical study of
  dormant bugs,'' in {\em 11th Working Conference on Mining Software
  Repositories (MSR)}, 2014.

\bibitem{ray16}
B.~Ray, V.~Hellendoorn, S.~Godhane, Z.~Tu, A.~Bacchelli, and P.~Devanbu, ``On
  the naturalness of buggy code,'' in {\em 38th International Conference on
  Software Engineering (ICSE)}, 2016.

\bibitem{Wang:ICSM:2019}
Y.~{Wang}, ``What motivate software engineers to refactor source code?
  evidences from professional developers,'' in {\em International Conference on
  Software Maintenance (ICSM)}, pp.~413--416, 2009.

\bibitem{Mahmoudi:2019:SANER}
M.~{Mahmoudi}, S.~{Nadi}, and N.~{Tsantalis}, ``Are refactorings to blame? an
  empirical study of refactorings in merge conflicts,'' in {\em 26th
  International Conference on Software Analysis, Evolution and Reengineering
  (SANER)}, pp.~151--162, 2019.

\bibitem{Lin:SANER:2019}
B.~{Lin}, C.~{Nagy}, G.~{Bavota}, and M.~{Lanza}, ``On the impact of
  refactoring operations on code naturalness,'' in {\em 26th International
  Conference on Software Analysis, Evolution and Reengineering (SANER)},
  pp.~594--598, 2019.

\bibitem{Kim:icse:2016}
J.~{Kim}, D.~{Batory}, D.~{Dig}, and M.~{Azanza}, ``Improving refactoring speed
  by 10x,'' in {\em 38th International Conference on Software Engineering
  (ICSE)}, pp.~1145--1156, 2016.

\bibitem{Szoke:SANER:2016}
G.~{Szóke}, C.~{Nagy}, R.~{Ferenc}, and T.~{Gyimóthy}, ``Designing and
  developing automated refactoring transformations: An experience report,'' in
  {\em 23rd International Conference on Software Analysis, Evolution, and
  Reengineering (SANER)}, pp.~693--697, 2016.

\bibitem{Bavota:JournalSystem:2015}
G.~Bavota, A.~D. Lucia, M.~D. Penta, R.~Oliveto, and F.~Palomba, ``An
  experimental investigation on the innate relationship between quality and
  refactoring,'' {\em Journal of Systems and Software}, vol.~107, no.~C,
  pp.~1--14, 2015.

\bibitem{Bavota:SCAM:2012}
G.~{Bavota}, B.~{De Carluccio}, A.~{De Lucia}, M.~{Di Penta}, R.~{Oliveto}, and
  O.~{Strollo}, ``When does a refactoring induce bugs? an empirical study,'' in
  {\em 12th International Working Conference on Source Code Analysis and
  Manipulation (SCAM)}, pp.~104--113, 2012.

\bibitem{johnson:acm2006}
D.~Dig and R.~Johnson, ``How do {APIs} evolve? a story of refactoring,'' in
  {\em 22nd International Conference on Software Maintenance (ICSM)},
  pp.~83--107, 2005.

\bibitem{Shen:OOPSLA:2019}
B.~Shen, W.~Zhang, H.~Zhao, G.~Liang, Z.~Jin, and Q.~Wang, ``{IntelliMerge}: A
  refactoring-aware software merging technique,'' {\em Programming Languages},
  vol.~3, no.~170, pp.~170:1--170:28, 2019.

\bibitem{jss-2018-jmove}
R.~Terra, M.~T. Valente, S.~Miranda, , and V.~Sales, ``{JMove}: A novel
  heuristic and tool to detect move method refactoring opportunities,'' {\em
  Journal of Systems and Software}, vol.~138, pp.~19--36, 2018.

\bibitem{Alves:2014:FSE}
E.~L.~G. Alves, M.~Song, and M.~Kim, ``{RefDistiller}: A refactoring aware code
  review tool for inspecting manual refactoring edits,'' in {\em 22nd
  International Symposium on Foundations of Software Engineering (FSE)},
  pp.~751--754, 2014.

\bibitem{Lin:2016:FSE}
Y.~Lin, X.~Peng, Y.~Cai, D.~Dig, D.~Zheng, and W.~Zhao, ``Interactive and
  guided architectural refactoring with search-based recommendation,'' in {\em
  24th International Symposium on Foundations of Software Engineering (FSE)},
  pp.~535--546, 2016.

\bibitem{Chaparro:ICSME:2014}
O.~{Chaparro}, G.~{Bavota}, A.~{Marcus}, and M.~D. {Penta}, ``On the impact of
  refactoring operations on code quality metrics,'' in {\em 30th International
  Conference on Software Maintenance and Evolution (ICSME)}, pp.~456--460,
  2014.

\bibitem{Fernandes:2019:ICSE}
E.~Fernandes, ``Stuck in the middle: Removing obstacles to new program features
  through batch refactoring,'' in {\em 41st International Conference on
  Software Engineering: Companion Proceedings (ICSE)}, pp.~206--209, 2019.

\bibitem{Tenorio:2019:IWOR}
D.~Tenorio, A.~C. Bibiano, and A.~Garcia, ``On the customization of batch
  refactoring,'' in {\em 3rd International Workshop on Refactoring (IWOR)},
  pp.~13--16, 2019.

\bibitem{Fernandes:2019:IWOR}
E.~Fernandes, A.~Uch\^{o}a, A.~C. Bibiano, and A.~Garcia, ``On the alternatives
  for composing batch refactoring,'' in {\em 3rd International Workshop on
  Refactoring (IWOR)}, pp.~9--12, 2019.

\bibitem{Jiau:TSE:2013}
H.~C. {Jiau}, L.~W. {Mar}, and J.~C. {Chen}, ``{OBEY}: Optimal batched
  refactoring plan execution for class responsibility redistribution,'' {\em
  Transactions on Software Engineering}, vol.~39, no.~9, pp.~1245--1263, 2013.

\bibitem{Meananeatra:ASE:2012}
P.~{Meananeatra}, ``Identifying refactoring sequences for improving software
  maintainability,'' in {\em 27th International Conference on Automated
  Software Engineering (ASE)}, pp.~406--409, 2012.

\end{thebibliography}

\end{document}